\documentclass[12pt]{article}
\usepackage{amssymb}
\usepackage{amsfonts}
\usepackage{youngtab}
\usepackage{amsmath,amssymb,latexsym,cite}
\usepackage{amsthm}
\usepackage{comment}
\usepackage{tikz-cd}
\usepackage{cite}
\usepackage{bbm}
\usepackage{xcolor}
\usepackage{amsthm}
\usepackage{xcolor}
\usepackage{circuitikz}
\usepackage{graphicx}
\usepackage{caption}
\usepackage{subcaption}

\newtheorem*{conj}{Conjecture}

\usepackage{hyperref}
\input xy
\xyoption{all}

\newcommand{\C}{\mathbb{C}}
\newcommand{\Z}{\mathbb{Z}}

\numberwithin{equation}{section}

\begin{document}

\vspace*{0.5in}

\begin{center}

{\large\bf Decomposition in 2d non-invertible gaugings}

Alonso Perez-Lona

\begin{tabular}{l}
Department of Physics, MC 0435\\
850 West Campus Drive\\
Virginia Tech\\
Blacksburg, VA  24061
\end{tabular}

\textit{Corresponding author: Alonso Perez-Lona,} {\tt aperezl@vt.edu}
$\,$

\end{center}

We extend the decomposition conjecture to 2d quantum field theories with a gauged $\text{Rep}(H)$ symmetry category for $H$ a finite-dimensional semisimple Hopf algebra with $\text{Rep}(G)$ trivially-acting and $\text{Vec}(\Gamma)$ the remaining symmetry, for $G,\Gamma$ finite groups. We check our extension by explicitly computing partition functions, and by verifying that previous results arise as special cases. Furthermore, we compute the topological operators responsible for enforcing the decomposition. Then, drawing from these results, we formulate a plausible decomposition conjecture for the even more general case of $\text{Rep}(H'')$ trivially-acting and $\text{Rep}(H')$ the remaining symmetry, for $H',H''$ Hopf algebras, not necessarily associated with groups.

\begin{flushleft}
January 2026
\end{flushleft}

\newpage

\tableofcontents

\newpage

\section{Introduction}

The connection between gauged trivially-acting finite groups, and decomposition in two dimensions, is by now well-known. On the one hand, decomposition, the splitting of a single theory into a disjoint sum of independent theories, was originally observed in the context of two-dimensional theories \cite{Hellerman:2006zs}, such as orbifolds, and gauge theories (see \cite{Sharpe:2022ene} for a streamlined review). Since then, decomposition has led to general statements about 2d (e.g. \cite{Robbins:2020msp,Robbins:2021ibx,Sharpe:2021srf,Nguyen:2021naa,Pantev:2023dim,Perez-Lona:2023djo,Bhardwaj:2024qrf,Perez-Lona:2024sds}), 3d (e.g. \cite{Pantev:2022kpl,Pantev:2022pbf,Perez-Lona:2023llv,Pantev:2024kva}), and 4d (e.g. \cite{Tanizaki:2019rbk}) quantum field theories. On the other hand, the idea of global symmetries has recently undergone a significant extension, resulting in what is nowadays known as generalized symmetries \cite{Gaiotto:2014kfa, Sharpe:2015mja} (see also the lecture notes \cite{Schafer-Nameki:2023jdn,Shao:2023gho,Bhardwaj:2023kri} and references therein), and in particular non-invertible symmetries. Thus, decomposition and non-invertible symmetries are general features of quantum field theories. The purpose of this paper is to systematically describe decomposition in two-dimensional theories with gauged non-invertible symmetries, in which a subsymmetry acts trivially.

In the context of (finite) non-invertible symmetries in two dimensions, symmetries are described in full generality by (multi-)fusion categories. While the additional layer of information, namely, morphisms between objects, makes the overall picture significantly richer, it also makes algebraic manipulations more convoluted. Finite-dimensional semisimple Hopf algebras provide a compromise between both pictures: while via their representation categories they give rise to very general fusion categories, their algebraic data, and in particular extension theory, is conceptually similar to that of groups.

In \cite{Hellerman:2006zs}, a concrete decomposition conjecture was proposed for 2d theories with finite group gauge symmetries with trivially-acting subsymmetries. Partition function computations provide robust evidence for the conjecture. In this paper, we propose an extension of this conjecture to non-invertible symmetries given by categories of representations of Hopf algebras, where the trivially-acting symmetry is of the form $\text{Rep}(G)$, and the remaining symmetry is of the form $\text{Vec}(\Gamma)$, for $G,\Gamma$ a pair of finite groups. We check this conjecture by computing partition functions of the decomposing theories, in agreement with concrete examples already present in the literature \cite{Perez-Lona:2023djo,Bhardwaj:2024qrf,Perez-Lona:2024sds}. Then, drawing from these results, we comment on a plausible decomposition conjecture for the even more general situation of $\text{Rep}(H'')$ trivially-acting, for $H''$ a Hopf algebra. 

The reader should note that all the fusion categories considered in this paper are representation categories of some finite-dimensional semisimple Hopf algebra and, as such, always admit a fiber functor, which physically is interpreted as a fully symmetry-preserving gapped phase \cite{Bhardwaj:2017xup,Thorngren:2019iar}. Not all fusion categories admit such a gapped phase, as exemplified by those of the form $\text{Vec}(G,\alpha)$ for $[\alpha]\in H^3(G,U(1))$ a nontrivial associator. It is, however, possible to interpret some fusion categories that do not admit said gapped phase as representation categories of \textit{quasi}-Hopf algebras \cite{Perez-Lona:2025ncg}. While we do not address this in the present paper, the arguments presented here at least conceptually extend to those cases. We leave a careful treatment of this for future work.

In Section~\ref{sec:general}, we state our conjecture. In Section~\ref{sec:mathbackground}, we heuristically introduce the necessary mathematical objects relevant for this paper, whose technical details are presented in Appendix~\ref{app:notes}. In Section~\ref{sec:pfuncdecomposition}, we concentrate on torus partition function computations for the special case of abelian extensions. We explicitly construct the special symmetric Frobenius algebra to gauge, and argue that the torus partition function splits in a way predicted by the general conjecture. Section \ref{sec:decomposition-operators} is concerned with the construction of the topological operators responsible for the decomposition. In Section~\ref{sec:examples}, we spell out some examples. Finally, in Section~\ref{sec:generalextensions}, drawing from observations both from the conjecture in \cite{Hellerman:2006zs} and from the results from Section~\ref{sec:pfuncdecomposition}, we describe a construction for a plausible decomposition conjecture for more general Hopf algebra extensions. In particular, we show how this general construction can recover previous statements. Appendix~\ref{app:sweedler} contains a brief reminder of Sweedler's notation for Hopf algebras, used mainly in Section~\ref{sec:generalextensions}.

\paragraph{Acknowledgments.} We would like to thank Eric Sharpe, Daniel Robbins, and Thomas Vandermeulen for discussions on the contents of Section~\ref{sec:decomposition-operators}.

\section{General conjecture}\label{sec:general}

Given a two-dimensional theory $\cal T$ with a finite group symmetry $\Gamma$, and a trivially-acting normal subgroup $N$, fitting in the exact sequence of groups
\begin{equation}\label{eq:general-gpsequence}
    1\to N\to \Gamma\to G\to 1,
\end{equation}
the decomposition conjecture \cite{Hellerman:2006zs} states that the $\Gamma$-gauged theory $[{\cal T}/\Gamma]$ decomposes as
\begin{equation}\label{eq:general-gpdecomp}
    [{\cal T}/\Gamma] = \bigoplus_{i\in \text{Irrep}(N)/G} [{\cal T}/K_i]_{\omega_i},
\end{equation}
where the sum runs over $\text{Irrep}(N)/G$, the set of orbits of the irreducible representations of $N$ by an action by $G$, $K_i\leq G$ is the stabilizer of the orbit $i$, and $\omega_i\in Z^2(K_i,U(1))$ is a discrete torsion 2-cocycle.

Non-invertible symmetries in two dimensions are specified by (multi-)fusion categories. Finite-dimensional semisimple Hopf algebras provide, through their representation categories, interesting yet manageable example of such categories. Therefore, it is natural to expect a decomposition conjecture for symmetries described by representation categories of Hopf algebras.

We specialize to particular kinds of finite-dimensional semisimple Hopf algebras known as \textit{abelian} extensions. These describe the case of a $\text{Rep}(H)$ symmetry with a $\text{Rep}(G)$-trivially acting subsymmetry and $\text{Vec}(\Gamma)$ the remaining symmetry, for $G,\Gamma$ a pair of groups. As one might guess, these categories fit in an \textit{exact} \textit{sequence} of fusion categories
\begin{equation}\label{eq:general-abeliansequence}
    \text{Rep}(G)\to \text{Rep}(H)\to \text{Vec}(\Gamma),
\end{equation}
defined in Section~\ref{sec:mathbackground}.

In this paper, we propose that the analogue of (\ref{eq:general-gpdecomp}) for the setting (\ref{eq:general-abeliansequence}) is
\begin{equation}\label{eq:general-abconjecture}
    [{\cal T}/\text{Rep}(H)] = \bigoplus_{i\in G/\Gamma} [{\cal T}/K_i]_{\omega_i},
\end{equation}
where the sum runs over $G/\Gamma$ the orbits of the \textit{set} $G$ by an action of $\Gamma$, $K_i\leq \Gamma$ is the stabilizer of the orbit $i$, and $\omega_i\in Z^2(K_i,U(1))$ is some particular discrete torsion. By $[{\cal T}/\text{Rep}(H)]$, we mean the theory resulting from gauging the symmetric special Frobenius algebra $(A,\mu,\Delta)$ determined by the fiber functor on $\text{Rep}(H)$ that reconstructs the Hopf algebra $H$.

An interesting observation of the construction of the relevant Frobenius algebra is that the Frobenius algebra is independent of half of the Hopf algebra extension information, namely the algebra structure of $H$. This explains the striking similarity of the conjectures (\ref{eq:general-gpdecomp}) and (\ref{eq:general-abconjecture}). Schematically, both consist of three different parts:
\begin{enumerate}
    \item An orbit set of the simples of $\text{Rep}((H'')^*)$, the representation category of the Hopf algebra $(H'')^*$ \textit{dual} to $H''$, where $\text{Rep}(H'')$ is the trivially-acting subcategory, by an action of the remaining symmetry category,
    \item for each orbit, a stabilizer subcategory of the remaining symmetry category,
    \item some discrete torsion that gauges the stabilizer subcategory.
\end{enumerate}
In \cite{Perez-Lona:2023djo}, it was shown that the Frobenius multiplication for the Frobenius algebra that gauges $\text{Rep}(H)$ is entirely determined by the comultiplication of $H$, and in \cite{Diatlyk:2023fwf} it was argued that the Frobenius comultiplication can be computed simply as the adjoint morphism to the Frobenius multiplication. Thus, it seems plausible that the decomposition of more general Hopf extensions is again independent of the algebra structure of $H$ and thus exhibits the same schematic form described above.

Assuming this is the case, then we propose that the decomposition of a $\text{Rep}(H)$-gauged theory with $\text{Rep}(H'')$ trivially-acting and $\text{Rep}(H')$ the remaining symmetry, fitting in the exact sequence
\begin{equation}
    \text{Rep}(H'')\to\text{Rep}(H)\to\text{Rep}(H'),
\end{equation}
takes the following form
\begin{equation}
    [{\cal T}/\text{Rep}(H)] = \bigoplus_{i\in \mathcal{O}} \, [{\cal T}/(K_i,\mu_i,\Delta_i)],
\end{equation}
where the sum runs over what may be understood as the orbits of the simples of $\text{Rep}((H'')^*)$ under the action of $\text{Rep}(H')$, $K_i$ is the regular object of the stabilizer subcategory of the orbit $i$, and $(\mu_i,\Delta_i)$ is some discrete torsion choice describing a possibly \textit{non-invertible} gauging (see  \cite{Perez-Lona:2023djo,Perez-Lona:2024sds,Diatlyk:2023fwf} for gauging symmetric special Frobenius algebras of non-invertible symmetries). We make all these components precise in Section~\ref{sec:generalextensions}.

\section{Mathematical background}\label{sec:mathbackground}

The main evidence we will present for the decomposition conjecture (\ref{eq:general-abconjecture}) is partition function computations. In this section, we give the intuition behind the mathematical objects and essential definitions needed for these computations. We cover this content in technical terms in Appendix~\ref{app:notes}. Throughout this paper, a \textit{subsymmetry} will be represented by a fusion subcategory\footnote{In contrast to just a symmetric special Frobenius algebra $A$ in $\cal C$, as the term is sometimes understood in the context of 2d non-invertible symmetries.} $\cal D\subset C$. 

\subsection{Exact sequences of fusion categories}\label{ssec:exactsequencesoffusion}

In the case of group-like symmetries, exact sequences are used to describe trivially-acting subsymmetries. If a theory $\cal T$ has a global symmetry parameterized by a finite group $\Gamma$ such that a normal subgroup $N \unlhd\Gamma$ acts trivially, then the $\Gamma$-action on $\cal T$ factors through its quotient group $G=\Gamma/N$, which is the remaining symmetry. This gives rise to the exact sequence
\begin{equation}\label{eq:gpexactseq}
    1\to N\xrightarrow{i} \Gamma \xrightarrow{\pi} G\to 1,
\end{equation}
for $i:N\hookrightarrow \Gamma$ the inclusion, and $\pi: \Gamma\to G$ the projection.

Thus, the scenario of a theory $\cal T$ with a global symmetry parameterized by a fusion category $\cal C$ such that a fusion subcategory $\cal K$ acts trivially should be similarly encoded by an \textit{exact sequence of fusion categories}
\begin{equation}
    {\cal K} \xrightarrow{\imath} {\cal C}\xrightarrow{F} {\cal D}.
\end{equation}
The intuitive picture to keep in mind is that $\cal K$ is a ``normal subcategory'' of $\cal C$, and that $\cal D$ is the "quotient" of $\cal C$ by $\cal K$. The functors $\imath$ and $F$ come equipped with a strong monoidal structure, which allows us to transport fusion rules and algebra objects consistently from one fusion category to another. See Appendix~\ref{sapp:exactsequencesoffusion} for the specific axioms.

\subsection{Exact sequences of Hopf algebras}\label{ssec:hopfexact}
Concrete examples of exact sequences of fusion categories are provided by exact sequences of Hopf algebras. Here, we will take all Hopf algebras to be more specifically finite-dimensional semisimple Hopf algebras. See e.g. \cite[Appendix 1]{Perez-Lona:2023djo} for the relevant axioms defining Hopf algebras.

An exact sequence, equivalently an \textit{extension}, of Hopf algebras is a diagram of Hopf algebras of the form
\begin{equation}\label{eq:1exactseqhopf}
    H' \xrightarrow{i} H \xrightarrow{\pi} H'', 
\end{equation}
where $i,\pi$ are Hopf algebra homomorphisms satisfying the conditions in Appendix~\ref{sapp:generalextensions}. As for fusion categories, the last object $H''$ is thought of as the quotient of $H$ by the Hopf subalgebra $H'$.

Applying the contravariant functor $\text{Rep}(-)$ to the sequence (\ref{eq:1exactseqhopf}) results in an exact sequence of fusion categories \cite{BN11}
\begin{equation}
    \text{Rep}(H'')\to \text{Rep}(H)\to \text{Rep}(H').
\end{equation}

Exact sequences of fusion categories and of Hopf algebras both specialize to the familiar group-like case. Indeed, given a sequence (\ref{eq:gpexactseq})
\begin{equation*}
    1\to N\to \Gamma\to G\to 1,
\end{equation*}
this defines an exact sequence of dual group algebras
\begin{equation}
    \C^{G}\to \C^{\Gamma}\to \C^{N},
\end{equation}
where $\C^G$ denotes the Hopf algebra of functions on $G$ valued in $\C$, and which in turn, by using $\text{Rep}(-)$, defines the sequence of fusion categories
\begin{equation}
    \text{Vec}(N)\to \text{Vec}(\Gamma)\to\text{Vec}(G).
\end{equation}
This sequence of fusion categories is what we understand as a $\Gamma$-symmetry with a trivially-acting $N$-symmetry and a remaining $G$-symmetry, all non-anomalous (in the sense that each admits a fiber functor/Symmetry Protected Topological (SPT) phase \cite{Choi:2023xjw,Bhardwaj:2024qrf}).

\

An accessible subclass of Hopf algebra extensions is given by imposing the additional assumptions $H''=\C G$ and $H'=\C^{\Gamma}$ for $G,\Gamma$ a pair of finite groups. Here, $\C G$ denotes the group Hopf algebra of $G$ over the complex numbers. Sequences of this form
\begin{equation}\label{eq:abextension}
    \C^{\Gamma}\to H\to \C G,
\end{equation}
are called \textbf{\textit{abelian} extensions}. 

All Hopf algebra extensions (\ref{eq:1exactseqhopf}) are classified by \textit{extension data}. As summarized in \cite{Nat20} (cf. \cite{Tak81,Mat02}), in the special case of abelian extensions, the data consists of:
\begin{itemize}
\item A \textit{matched pair} \cite{Kac68} of groups $(G,\Gamma,\triangleright,\triangleleft)$: a pair of finite groups $G,\Gamma$ along with maps of sets
\begin{eqnarray}
    \triangleleft:& \Gamma\times G\to \Gamma,
    \\
    \triangleright:& \Gamma\times G \to G, \label{eq:gammaactong}
\end{eqnarray}
satisfying the axioms (\ref{eq:trianglerightidentity})-(\ref{eq:triangleleftifentity}) of actions by permutation,

\item and a pair of 2-cocycles $\sigma: G\times G\to (\C^{\times})^{\Gamma}$, $\tau: \Gamma\times\Gamma\to (\C^{\times})^G$ satisfying the normalized 2-cocycle conditions (\ref{eq:sigma2-cocycle})-(\ref{eq:2cocyclesevaluatedatone}), where $(\C^{\times})^G$ means the convolution-invertible complex-valued functions on $G$.

We will use the notation $\sigma_s(g,h)=\sigma(g,h)(s)$ (and similarly for $\tau_g(t,u)$), for $g,h,l\in G$ and $s,t,u\in \Gamma$.
\end{itemize}
The advantage of this setting is that all extension data is formulated in terms of actions and cocycles, thus allowing us to bypass explicitly dealing with dual versions of these even though both the algebra and coalgebra structures can be nontrivial.

The Hopf algebra $H$ has an underlying vector space $\C^{\Gamma}\otimes \C G$, whose basis elements are customarily presented as $v_g\# x$ for $x\in G$ the usual basis of a group algebra, and $v_g$ the dual basis of the basis element given by $g\in \Gamma$.
The Hopf algebra structure $(u:\C\to H, \mu:H\otimes H\to H, \epsilon: H\to \C, \Delta: H\to H\otimes H)$ is given by \cite{AD95,Mat02}
\begin{gather}
    u: 1\mapsto \sum_{g\in\Gamma} v_g\otimes 1,\label{eq:abextunit}\\
\mu: (v_g\# x)\otimes (v_h\# y) \mapsto  \delta_{g\triangleleft x,h}\sigma_g(x,y)v_g\# xy, \label{eq:abextmult} \\
    \epsilon: (v_g\# x) \mapsto \delta_{1,g}, \label{eq:abextcounit}\\
    \Delta: (v_g\# x) \mapsto \sum_{t\in \Gamma} (\tau_x(gt^{-1},t)v_{gt^{-1}}\# t\triangleright x)\otimes (v_t\# x).\label{eq:abextcomult}
\end{gather}

The antipode is given by \cite[Lemma 2.3]{AN03}
\begin{equation}\label{eq:abantipode}
    S(v_g\# x) = \big(\sigma_{(g \triangleleft x)^{-1}}((g\triangleright x)^{-1}, g\triangleright x)\big)^{-1} \, \big(\tau_x(g^{-1}, g)\big)^{-1}\, v_{(g\triangleleft x)^{-1}} \# (g\triangleright x)^{-1}.
\end{equation}

Finally, as a finite-dimensional Hopf algebra, it has an integral, and cointegral. As verified in Appendix~\ref{sapp:abelian}, the integral $\Lambda\in H$ is
\begin{equation}\label{eq:hopfintegral}
    \Lambda = v_1\# \left(\sum_{x\in G}x\right).
\end{equation}
A cointegral $\lambda$ of $H$ is the function
\begin{equation}
    \lambda=\left(\sum_{g\in\Gamma} g\right)\#v_1.
\end{equation}
This data will be used to construct the symmetric Frobenius algebra that gauges $\text{Rep}(H)$.

\section{Consistency check: partition function computations}\label{sec:pfuncdecomposition}

In this section, we compute genus-one partition functions for gauged abelian Hopf extensions, in order to check the decomposition conjecture (\ref{eq:general-abconjecture}).  This involves constructing the relevant symmetric special Frobenius (or, for short, gaugeable) algebra, as well as suitable monoidal functors to relate the partition function of the theory to those of its constituent universes.

The exact sequence one obtains from an abelian extension (\ref{eq:abextension}) of Hopf algebras is of the form
\begin{equation}\label{eq:abextcategorysequence}
    \text{Rep}(G)\to \text{Rep}(H)\xrightarrow{F} \text{Vec}(\Gamma),
\end{equation}
where $F$ is the functor induced by the inclusion of Hopf algebras $i: \C^{\Gamma}\to H$ which on basis elements is
\begin{eqnarray}
    i:& \C^{\Gamma}&\to H,
    \\
    & v_g&\mapsto v_g\# 1.
\end{eqnarray}

Decomposition in this case is easier to describe, as the gaugeable algebras $\text{Vec}(\Gamma)$ with a single copy of the monoidal unit $\mathbbm{1}_{\text{Vec}(\Gamma)}=1$ correspond to pairs $(H,\omega)$ where $H\leq \Gamma$ is a subgroup of $\Gamma$, and $[\omega]\in H^2(H,U(1))$ is the cohomology class of a group 2-cocycle \cite{Ostrik:2002ohv}. In terms of this classification, the algebra appearing here corresponds to the pair $(\Gamma,1)$, interpreted as gauging the whole $\Gamma$-symmetry with no discrete torsion.

\subsection{Gaugeable algebra}\label{ssec:frobeniusalgebra}

The first step is to construct a gaugeable algebra in $\text{Rep}(H)$. We remind the reader that this algebra can be used to gauge on a surface of \textit{any} genus. We will use the algebra based on the dual Hopf algebra $H^*$, following the process in \cite{Perez-Lona:2023djo}. This is the vector space $H^*=\C\Gamma\otimes\C^G$, with basis $g\otimes e_x$ for $g\in\Gamma, x\in G$. The comultiplication $\Delta:H\to H\otimes H$ in (\ref{eq:abextcomult}) gives the multiplication $m:H^*\otimes H^*\to H^*$ on $H^*$ by dualizing
\begin{eqnarray}
    \mu_F:=\Delta^*:& H^*\otimes H^*&\to H^*, \label{eq:frobeniusmultisadual}
    \\
    & (g\# v_x)\otimes (h\# v_y) &\mapsto \tau_y(g,h) gh\# \delta_{x,h\triangleright y}v_y.
\end{eqnarray}
The unit $u_F:\C\to H^*$ is the dual map of the counit (\ref{eq:abextcounit})
\begin{equation}\label{eq:frobenius-unit-dual}
    u_F = \sum_{x\in G} 1\# v_x.
\end{equation}
The general formula for the comultiplication \cite[Eq. 2.57]{Perez-Lona:2023djo} (cf. \cite{FSS11}) is
\begin{equation}\label{eq:deffrobeniuscomult}
    \Delta_F := c\,(\text{Id}_H \otimes (\lambda \circ \mu)) \circ (\text{Id}_H \otimes S \otimes  \text{Id}_H) \circ (\Delta \otimes  \text{Id}_H))^*,
\end{equation}
for $c\in \C^{\times}$ a normalization constant. The morphism to dualize is defined as
\begin{equation}
    m:=\text{Id}_H \otimes (\lambda \circ \mu)) \circ (\text{Id}_H \otimes S \otimes  \text{Id}_H) \circ (\Delta \otimes  \text{Id}_H).
\end{equation}
Using this, we derive that
\begin{gather}
    m((v_g\# x)\otimes (v_h\# y)) = \sum_{t\in\Gamma} \delta_{t, (h\triangleleft y)^{-1}\triangleleft x^{-1}} \, \delta_{t\triangleright x,y}  (\tau_x(g,((h\triangleleft y)^{-1}\triangleleft x^{-1})^{-1}))^{-1}
    v_{g((h\triangleleft y)^{-1}\triangleleft x^{-1})^{-1}}\# y,
\end{gather}
so that the Frobenius comultiplication is given by
\begin{equation}
    \Delta_F(k\# v_z)=c\,\sum_{g \in\Gamma} \big( \tau_{g^{-1}k \triangleright z}(g,g^{-1}k)\big)^{-1}\,
    \big( (g\# v_{g^{-1}k \triangleright z})\otimes (g^{-1}k\# v_z)\big).
\end{equation}
Finally, the counit $\epsilon_F: H^*\to \C$ is given by the dual of the integral $\Lambda$ (\ref{eq:hopfintegral}):
\begin{equation}
    \epsilon_F(g\# v_x) = (g\# v_x)(\Lambda) = v_1(g) v_x\left(\sum_{y\in G} y\right) = \delta_{1,g}.
\end{equation}

Therefore, the symmetric special Frobenius algebra $(H^*,\mu_F,u_F,\Delta_F,\epsilon_F)$ that gauges the fusion category $\text{Rep}(H)$ is given by
\begin{equation}
        u_F(1) = \sum_{x\in G} 1\# v_x, \label{eq:frobeniusunit} 
\end{equation}
\begin{equation}
    \mu_F(g\# v_x,h\# v_y) = \tau_y(g,h) gh\# \delta_{x,h\triangleright y}v_y, \label{eq:frobeniusmult} 
\end{equation}
\begin{equation}
       \epsilon_F(k\# v_z) =  \delta_{1,k}, \label{eq:frobeniuscounit}
\end{equation}
\begin{equation}
    \Delta_F(k\# v_z) =  \frac{1}{\vert\Gamma\vert} \sum_{g \in\Gamma} \big( \tau_{g^{-1}k \triangleright z}(g,g^{-1}k)\big)^{-1}\,
    \big( (g\# v_{g^{-1}k \triangleright z})\otimes (g^{-1}k\# v_z)\big) \label{eq:frobeniuscomult},
\end{equation}
which is the canonical gauging in the sense that it comes from the fiber functor of $\text{Rep}(H)$ that via Tannaka reconstruction \cite{EGNO} recovers $H$. Demanding the special condition $\mu_F\circ\Delta_F=\text{id}$ gives the normalization constant $c=\tfrac{1}{\vert\Gamma\vert}$.

There is a crucial fact to highlight here. Observe that the algebra extension information, which consists of the 2-cocycle $\sigma: G\times G\to (\C^{\times})^{\Gamma}$ and the action by permutation $\Gamma\times G\xrightarrow{\triangleleft} \Gamma$, does \textit{not} play a role in the gaugeable algebra structure of $H^*$, namely its discrete torsion \textit{choice} (cf. \cite{Perez-Lona:2024yih}). Naively, one would expect the 2-cocycle $\sigma$ to appear given that the Frobenius comultiplication (\ref{eq:deffrobeniuscomult}) is constructed from both the Hopf multiplication (\ref{eq:abextmult}) and comultiplication (\ref{eq:abextcomult}). However, the inclusion of the antipode $S$ in the computation precisely cancels out the $\sigma$-cocycle. 

The fact that the algebra extension information does not play a role in defining the discrete torsion of the regular $H$-module already had appeared in the literature. In \cite{Diatlyk:2023fwf}, the comultiplication comes from taking the adjoint of the multiplication morphism $\mu_F:H^*\otimes H^*\to H^*$. However, the Frobenius multiplication (\ref{eq:frobeniusmultisadual}) is constructed only from the coalgebra information of $H$. 

\subsection{Partition function computation}
In the finite group symmetry decomposition conjecture \cite{Hellerman:2006zs}, the partition function decomposition is obtained by first computing the partition function of the non-effectively acting symmetry $\Gamma$, and then quotienting out the trivially-acting part using the projection group homomorphism $\pi: \Gamma\to G$ to the remaining finite symmetry group. Intuitively, this looks like
\begin{equation}
    Z_{g,h}\mapsto Z_{\pi(g),\pi(h)}.
\end{equation}
In \cite{Perez-Lona:2024yih,Perez-Lona:2024sds}, partition function decompositions for a projection of symmetries $\rho:\text{Rep}(\Gamma)\to\text{Rep}(G)$ were obtained by first computing the partition function in $\text{Rep}(\Gamma)$, and then transferring the partition function to the remaining symmetry category $\text{Rep}(G)$, in analogy to the scenario above. This was done at a schematic level given the complexity of constructing the strong monoidal structure of the functor $\rho$.

Since we are only interested in observing how the $\text{Rep}(H)$-gauged partition function looks like in terms of $\text{Vec}(\Gamma)$-orbifolds, a different approach is to first transfer the gaugeable algebra $(H^*,\mu_F,\Delta_F)$ to $\text{Vec}(\Gamma)$ and only then gauge it in $\text{Vec}(\Gamma)$. This has the advantage that we do not need to deal with F-symbols in $\text{Vec}(\Gamma)$, as all the associator information can be taken to be trivial in said category. One can proceed this way because the restriction functor $F:\text{Rep}(H)\to \text{Vec}(\Gamma)$ is a tensor functor, so that it preserves gaugeable algebras. Therefore, one may talk about the gaugeable algebra $F((H^*,\mu_F,\Delta_F))$ in $\text{Vec}(\Gamma)$. For simplicity, we will denote this algebra as $(A,\mu_F,\Delta_F)$.

\subsubsection{Module decomposition}

As a first step, we want to obtain the decomposition of the object $A:=F(H^*)\in\text{Vec}(\Gamma)$ into simples
\begin{equation}\label{eq:simpledecomp}
    A = \sum_{g\in\Gamma} N^g_{A} U_g,
\end{equation}
for $U_g$ the 1-dimensional $\Gamma$-graded homogeneous vector space with degree $g\in \Gamma$. That is, we exhibit the $\Gamma$-grading ($\C\Gamma$-comodule structure) on $H^*$, which is given by composing the dual of the multiplication (\ref{eq:abextmult})
\begin{eqnarray}
    \mu^*:& H^*&\to H^*\otimes H^*
    \\
    & g\# v_x &\mapsto \sum_{y\in G} \sigma_{g}(y, y^{-1}x)(g\# v_y)\otimes ((g\triangleleft y)\# v_{ y^{-1}x}), \label{eq:module-decompo-coprod}
\end{eqnarray}
with the dual inclusion $i^*: g\# v_y \mapsto g \,\delta_{1,y}$,
\begin{eqnarray}\label{eq:mod-decomp-circ}
    (i^*\otimes\text{id}_{H^*})\circ \mu^*: g\# v_x \mapsto g\otimes (g\# v_x).
\end{eqnarray}
As one would expect, this says that in the decomposition (\ref{eq:simpledecomp}), $N^g_{A}=\vert G\vert$ for all $g\in \Gamma$, meaning $F(H^*)$ consists of $\vert G\vert$ copies of the regular object in $\text{Vec}(\Gamma)$.

\subsubsection{Torus partition function}

We now proceed to explicitly compute the genus one partition function for the Frobenius algebra $(A,\mu_F,\Delta_F)\in \text{Vec}(\Gamma)$ constructed in Section~\ref{ssec:frobeniusalgebra}.

As explained in \cite{Perez-Lona:2023djo}, the genus $1$ partition function for a Frobenius algebra $(A,\mu,\Delta)$ in a fusion category that is multiplicity-free\footnote{Namely, that $\text{dim}(\text{Hom}(L_1\otimes L_2,L_3))\leq 1$ for all $L_1,L_2,L_3$ simple objects in the fusion category.}, as is the case for $\text{Vec}(\Gamma)$, is computed as
\begin{equation}
    Z_A = \sum_{L_1,L_2,L_3} \mu_{L_1,L_2}^{L_3} \Delta_{L_3}^{L_2,L_1}\, Z_{L_1,L_2}^{L_3},
\end{equation}
for $L_1,L_2,L_3$ running over the simple objects of the fusion category. The coefficients are obtained as follows. We first expand the morphisms $\mu\in\text{Hom}(A\otimes A,A)$ and $\Delta\in\text{Hom}(A,A\otimes A)$ in terms of mappings of simple objects \cite[Eqn's 2.91-93]{Perez-Lona:2023djo}
\begin{equation}
    \mu_{L_1,L_2}^{L_3}: \text{Hom}(L_1,A)\otimes \text{Hom}(L_2,A)\to \text{Hom}(A,L_3),
\end{equation}
\begin{equation}
    \Delta_{L_3}^{L_2,L_1}: \text{Hom}(L_3,A)\to \text{Hom}(A,L_2)\otimes \text{Hom}(A,L_1).
\end{equation}
Then, we sum over the possible embeddings in $\text{Hom}(L_1,A)$, 
 $\text{Hom}(L_2,A)$, $\text{Hom}(A,L_3)$, to obtain the scalars $\mu_{L_1,L_2}^{L_3}\in \text{Hom}(L_1\otimes L_2,L_3)$, $\Delta_{L_3}^{L_2,L_1}\in \text{Hom}(L_3,L_2\otimes L_1)$.

Since in the present case
\begin{equation}
    A = \bigoplus_{x\in G} \left(\bigoplus_{g\in \Gamma} U_g\right),
\end{equation}
there are $\vert G\vert $ linearly-independent ways of embedding any simple object $U_g\in \text{ob}(\text{Vec}(\Gamma))$, including the monoidal unit $U_1$, into $A$. This means that each Hom-space $\text{Hom}(U_g,A)$ is $|G|$-dimensional. We choose a basis for the Hom-space $\text{Hom}(U_g,A)$ labeled by $x\in G$, and a basis for $\text{Hom}(U_h,A)$ by $y\in G$. This is because the embeddings are explicitly $x: g\mapsto g\# v_x$ and $y: h\mapsto h\# v_y$. Moreover, the fusion rules of $\text{Vec}(\Gamma)$ imply that $\text{dim}(\text{Hom}(U_g \otimes U_h,U_k))= \delta_{gh,k}$, meaning that only the partial traces $Z_{g,h}:=Z_{g,h}^{gh}$ can have non-zero coefficients. The embedding of $gh$ is also uniquely specified by the Frobenius multiplication (\ref{eq:frobeniusmult}) in terms of the embedding $x$ of $g$ and $y$ of $h$.

Taking this into account, the partition function is computed as
\begin{equation}
    Z_{(A,\mu_F,\Delta_F)} = \sum_{x,y\in G} \sum_{g,h\in \Gamma} \left((\Delta_F)_{gh}^{h,g}\,(\mu_F)_{g,h}^{gh}\right)_{x,y} \, Z_{g,h},
\end{equation}
where the $\vert G \vert$ linearly-independent ways to embed each simple object in $A$ appear as the sum over the embeddings $x,y\in G$.

We now expand the morphism $\Delta_F\circ\mu \in \text{Hom}(A\otimes A, A\otimes A)$:
\begin{eqnarray}
    \lefteqn{(\Delta_F\circ\mu_F)((g\# v_x)\otimes (h\# v_y))} 
    \\
    &=&  {\frac{1}{\vert\Gamma\vert} \sum_{t\in\Gamma} \tau_y(g,h) \,\delta_{x,h\triangleright y} \, \big(\tau_{t^{-1}gh\triangleright y} (t,t^{-1}gh) \big)^{-1} \, \big((t\# v_{t^{-1}gh\triangleright y})\otimes (t^{-1}gh\# v_y) \big).} \nonumber
\end{eqnarray}
From this we can read that
\begin{equation}
 \big((\Delta_F)_{gh}^{h,g}\,(\mu_F)_{g,h}^{gh}\big)_{x,y} = \frac{1}{\vert\Gamma\vert} \, \delta_{x,y} \, \delta_{ghg^{-1},h} \, \tau_y(g,h) \, \delta_{x,h\triangleright y} \, \big(\tau_{g\triangleright y} (h,g) \big)^{-1}.
\end{equation}
This gives the partition function
\begin{equation}
    Z_{(A,\mu_F,\Delta_F)} = \frac{1}{\vert\Gamma \vert} \sum_{x\in G} \left(\sum_{g,h\in \Gamma}  \delta_{gh,hg} \, \delta_{x,g\triangleright x} \, \delta_{x,h\triangleright x}  \frac{\tau_x(g,h)}{\tau_{x}(h,g)} \,Z_{g,h}\right).
\end{equation}
Thus, for a fixed group element $x\in G$, one sums over commuting pairs $(g,h)\in \Gamma\times \Gamma$ that act trivially on $x$ via the action by permutation. Denoting the stabilizer of $x$ as $K_x$, we can rewrite the partition function as
\begin{equation}\label{eq:beforepartfunct}
     Z_{(A,\mu_F,\Delta_F)} = \frac{1}{\vert\Gamma\vert} \sum_{x\in G} \vert K_x \vert \left(\frac{1}{\vert K_x \vert}\sum_{g,h\in K_x}  \delta_{gh,hg} \frac{\tau_x(g,h)}{\tau_{x}(h,g)} \,Z_{g,h}\right).
\end{equation}

Now, suppose $y=k\triangleright x$ for some $k\in \Gamma$. Using the 2-cocycle identity (\ref{eq:tau2cocycle}) we see that
\begin{equation}
    \frac{\tau_y(kgk^{-1},khk^{-1})}{\tau_y(khk^{-1},kgk^{-1})}= \frac{\tau_x(g,h)}{\tau_x(h,g)}.
\end{equation}
Furthermore, a $(g,h)$-twisted sector and its conjugation by some $k\in \Gamma$ are equal
\begin{equation}
    Z_{g,h} = Z_{kgk^{-1},khk^{-1}}.
\end{equation}
One quick way\footnote{We thank Daniel Robbins for clarifying this point.} to see this is from a path integral perspective. In this setting, $Z_{g,h}$ consists of fields $\phi$ with boundary conditions
\begin{eqnarray}
    \phi(z+1) &=& h\cdot \phi(z),
    \\
    \phi(z+\tau) &=& g\cdot \phi(z),
\end{eqnarray}
for $\tau$ the torus modulus. The equivalence comes about simply by a change of variables $\phi \mapsto k\cdot \phi$ in the path integral.
This means that, for a given $x\in G$, the terms inside the bracket in (\ref{eq:beforepartfunct}) are identical for every element $y\in 
\mathcal{O}_{\Gamma}(x)$ in the $\Gamma$-orbit of $x$. 

Finally, we use the orbit-counting theorem \cite[Chapter 3]{Rot12}. For $S$ a $\Gamma$-set, the theorem states that
\begin{equation}
    \sum_{[s]\in S/\Gamma}1 = \frac{1}{\vert \Gamma \vert} \sum_{s\in S} \vert \text{Stab}_{\Gamma}(s)\vert,
\end{equation}
where the left sum is simply a rewrite of $\vert S/\Gamma\vert$, and $\text{Stab}_{\Gamma}(s)$ is the stabilizer of $s\in S$, the subgroup of $\Gamma$ that fixes $s$. This result allows us to rewrite the partition function as
\begin{equation}\label{eq:2toruspartitionfunction}
    Z_{(A,\mu_F,\Delta_F)} = \sum_{[x]\in G/\Gamma} \left(\frac{1}{\vert K_x \vert} \sum_{g,h\in K_x} \frac{\tau_x(g,h)}{\tau_x(h,g)} \, Z_{g,h} \right) = \sum_{[x]\in G/\Gamma} Z_{K_x,\tau_x},
\end{equation}
for $G/\Gamma$ the set of $\Gamma$-orbits of $G$, and $Z_{K_x,\tau_x}$ the partition function of a $K_x\leq \Gamma$ orbifold with discrete torsion $\tau_x \in Z^2(K_x,U(1))$ turned on. As shown above, this is independent of the particular representative $x\in G$ of a given orbit. This partition function is consistent with the decomposition conjecture (\ref{eq:general-abconjecture}).

\section{Topological point operators and projectors}\label{sec:decomposition-operators}

So far, we have obtained evidence for the decomposition based on observing how the torus partition function splits. To establish the decomposition on firmer grounds, one ought to describe the dimension zero projection operators responsible for decomposition\footnote{We thank the anonymous referee of the Journal of Mathematical Physics (JMP) for suggesting this idea.}. 

These operators have been constructed for the case of invertible symmetries with trivially-acting normal subgroups in \cite[Section 2]{Sharpe:2021srf}. They have also been described in \cite{Robbins:2022wlr}, both for the same invertible setting as well as for the case of a non-invertible symmetry ${\rm Rep}(S_3)$, the whole of which acts trivially, matching the results in \cite[Section 4.6.1]{Bhardwaj:2023idu}. In Section~\ref{ssec:tpo-invertible}, we quickly review the main physical arguments of \cite{Robbins:2022wlr} relevant for this construction. Then, in Section~\ref{ssec:projection-operators}, by rephrasing these arguments in more algebraic terms, we construct the decomposition projection operators for our symmetry setup~(\ref{eq:general-abeliansequence}).

\subsection{Review of the invertible case}\label{ssec:tpo-invertible}

Consider a theory $\cal T$ with a global finite group symmetry $\Gamma$ and a trivially-acting normal subgroup $K \leq \Gamma$, which we take to be abelian for simplicity. As previously explained, in terms of symmetry fusion categories, this corresponds to the exact sequence of fusion categories
\begin{equation}
    {\rm Vec}(K) \xrightarrow{\imath} {\rm Vec}(\Gamma) \xrightarrow{F} {\rm Vec}(G)
\end{equation}
for $G:= \Gamma / K$. Because the Topological Defect Lines (TDL's) associated with the simples $L_k \in {\rm ob}({\rm Vec}(K))$ act trivially, they may each be joined to the trivial TDL $L_1$ via a point operator $\sigma_k$, a Topological Point Operator (TPO), as shown in Figure \ref{diagram:tpo-definition} (cf. \cite[Figure 4]{Robbins:2022wlr}). Up to rescaling, for each simple $L_k$ one gets exactly one TPO $\sigma_k$ because the latter lives in the junction hom-vector space $\sigma_k\in {\rm Hom}(\mathbbm{1},F(\imath(L_k)))$, which is one-dimensional since $L_k$ is invertible.

\begin{figure}[!ht]
\centering
\resizebox{0.4\textwidth}{!}{%
\begin{circuitikz}
\tikzstyle{every node}=[font=\LARGE]
\draw  (7.25,10.75) circle (0cm);

\node at (6.25,10.5) [circ] {};
\node [font=\large] at (5.25,11) {$L_1$};
\node [font=\large] at (7.5,11) {$L_k$};
\node [font=\large] at (6.25,10) {$\sigma_k$};
\draw [->, >=Stealth] (3.75,10.5) -- (5,10.5);
\draw [->, >=Stealth] (6.25,10.5) -- (7.5,10.5);
\draw [short] (5,10.5) -- (6.25,10.5);
\draw [short] (7.5,10.5) -- (8.75,10.5);
\end{circuitikz}
}%
\caption{A trivially-acting Topological Defect Line (TDL) $L_k$ is connected to the trivial TDL $L_1$ by a Topological Point Operator (TPO) $\sigma_k$.}\label{diagram:tpo-definition}
\end{figure}

Given that the trivial line $L_1$ is always invisible, it may be omitted in the diagrams, so that we may equally represent a TPO $\sigma_k$ as an operator living at the endpoint of a semi-infinite TDL $L_k$, as in Figure~\ref{diagram:tpo-definition2}.
\begin{figure}[!ht]
\centering
\resizebox{0.25\textwidth}{!}{%
\begin{circuitikz}
\tikzstyle{every node}=[font=\LARGE]
\draw  (7.25,10.75) circle (0cm);

\node at (6.25,10.5) [circ] {};
\node [font=\large] at (7.5,11) {$L_k$};
\node [font=\large] at (6.25,10) {$\sigma_k$};
\draw [->, >=Stealth] (6.25,10.5) -- (7.5,10.5);
\draw [short] (7.5,10.5) -- (8.75,10.5);
\end{circuitikz}
}%
\caption{The TPO $\sigma_k$ from Figure~\ref{diagram:tpo-definition} is equivalently depicted as an operator living at the end of a semi-infinite trivially-acting TDL $L_k$.}\label{diagram:tpo-definition2}
\end{figure}

These TPO's furthermore come equipped with an action by the group $\Gamma$, diagrammatically defined as in Figure~\ref{fig:tdl-action}. This action is given by the adjoint action of $\Gamma$ on $K$, which is well-defined since $K \trianglelefteq \Gamma$ is normal.
\begin{figure}[!ht]
\centering
\begin{minipage}{0.4\textwidth}
\centering
\begin{circuitikz}
\tikzstyle{every node}=[font=\large]
\node at (7.5,8.5) [circ] {};
\draw  (7.5,8.5) circle (1.25cm);
\draw [->, >=Stealth] (7.5,8.5) -- (7.5,9.25);
\draw [->, >=Stealth] (7.5,9.75) -- (7.5,10.5);
\draw [short] (7.5,9) -- (7.5,9.75);
\draw [short] (7.5,10.5) -- (7.5,11);
\draw [short] (7.75,9.75) -- (7.75,9.75);
\draw [->, >=Stealth] (7.5,7.25) -- (7.25,7.25);
\node [font=\large] at (8,9) {$L_k$};
\node [font=\large] at (7.5,8) {$\sigma_k$};
\node [font=\large] at (8.75,10.5) {$L_{{\rm ad}_{\gamma}(k)}$};
\node [font=\large] at (7.5,6.75) {$L_{\gamma}$};
\end{circuitikz}
\end{minipage}
\begin{minipage}{0.4\textwidth}
\centering
\begin{circuitikz}
\tikzstyle{every node}=[font=\LARGE]
\node at (7.5,8.5) [circ] {};
\draw [short] (7.75,9.75) -- (7.75,9.75);
\node [font=\large] at (7.5,8) {$\sigma_{{\rm ad}_{\gamma}(k)}$};
\node [font=\large] at (8.75,9.5) {$L_{{\rm ad}_{\gamma}(k)}$};
\draw [->, >=Stealth] (7.5,8.5) -- (7.5,9.75);
\draw [short] (7.5,9.75) -- (7.5,11);
\end{circuitikz}
\end{minipage}
\caption{Action of TDL $L_{\gamma}$ on TPO $\sigma_k$.}\label{fig:tdl-action}
\end{figure}

The adjoint action of $\Gamma$ on $K$ is well-known (e.g. \cite{Rot12}) to implement the action of $\triangleright:G\to {\rm Aut}(K)$ that characterizes, along with a cohomology class in $H^2(G,K)$, the group $\Gamma$ as a group extension
\begin{equation}\label{eq:tdl-gp-extension-review}
    1 \to K \to \Gamma \to G \to 1,
\end{equation}
by choosing any set-theoretic section $s: G \to \Gamma$. Therefore, one only needs to look at the action of the simple lines $L_g \in {\rm Vec}(G)$ in the effectively-acting fusion category, as depicted in Figure~\ref{fig:effective-tdl-action}.

\begin{figure}[!ht]
\centering
\begin{minipage}{0.4\textwidth}
\centering
\begin{circuitikz}
\tikzstyle{every node}=[font=\large]
\node at (7.5,8.5) [circ] {};
\draw  (7.5,8.5) circle (1.25cm);
\draw [->, >=Stealth] (7.5,8.5) -- (7.5,9.25);
\draw [->, >=Stealth] (7.5,9.75) -- (7.5,10.5);
\draw [short] (7.5,9) -- (7.5,9.75);
\draw [short] (7.5,10.5) -- (7.5,11);
\draw [short] (7.75,9.75) -- (7.75,9.75);
\draw [->, >=Stealth] (7.5,7.25) -- (7.25,7.25);
\node [font=\large] at (8,9) {$L_k$};
\node [font=\large] at (7.5,8) {$\sigma_k$};
\node [font=\large] at (8.75,10.5) {$L_{g \triangleright k}$};
\node [font=\large] at (7.5,6.75) {$L_{g}$};
\end{circuitikz}
\end{minipage}
\begin{minipage}{0.4\textwidth}
\centering
\begin{circuitikz}
\tikzstyle{every node}=[font=\large]
\node at (7.5,8.5) [circ] {};
\draw [short] (7.75,9.75) -- (7.75,9.75);
\node [font=\large] at (7.5,8) {$\sigma_{g \triangleright k}$};
\node [font=\large] at (8.75,9.5) {$L_{g \triangleright k}$};
\draw [->, >=Stealth] (7.5,8.5) -- (7.5,9.75);
\draw [short] (7.5,9.75) -- (7.5,11);
\end{circuitikz}
\end{minipage}
\caption{Action of the effectively-acting TDL $L_{g}\in {\rm ob}({\rm Vec}(G))$ on TPO $\sigma_k$.}\label{fig:effective-tdl-action}
\end{figure}

Now, as previously explained, to gauge the symmetry category ${\rm Vec}(\Gamma)$, one endows the regular object
\begin{equation}
    H = \bigoplus_{\gamma \in \Gamma}L_{\gamma}
\end{equation}
with the structure of a symmetric special Frobenius algebra $(H,\mu,\Delta)$ (a choice of discrete torsion). This regular object includes, in particular, the regular object
\begin{equation}
    R= \bigoplus_{k\in K} L_k
\end{equation}
of the trivially-acting category ${\rm Vec}(K)$. Notice that, since $K\leq \Gamma$, the tuple $(R,\mu\vert_R,\Delta\vert_R)$ is itself a special symmetric Frobenius algebra in ${\rm Vec}(\Gamma)$. The TPO's $\{\sigma_k\}_{k\in K}$ thus inherit a fusion structure, as shown in Figure~\ref{fig:tpo-fusion} (cf. \cite[Figure 11]{Robbins:2022wlr}).
\begin{figure}[h]
    \centering
    \begin{tabular}{ccc}
   \begin{circuitikz}
\tikzstyle{every node}=[font=\large]
\node at (5,10.5) [circ] {};
\node at (6.25,9.25) [circ] {};
\node at (7.5,10.5) [circ] {};
\draw [->, >=Stealth] (5,10.5) -- (4.25,11.25);
\draw [->, >=Stealth] (7.5,10.5) -- (8.25,11.25);
\draw [short] (4.25,11.25) -- (3.75,11.75);
\draw [short] (8.25,11.25) -- (8.75,11.75);
\node [font=\large] at (5,10.25) {$\sigma_{k_1}$};
\node [font=\large] at (7.5,10.25) {$\sigma_{k_2}$};
\node [font=\large] at (7.5,9) {$\sigma_{(k_1k_2)^{-1}}$};
\draw [->, >=Stealth] (6.25,8) -- (6.25,8.75);
\draw [short] (6.25,8.75) -- (6.25,9.25);
\node [font=\large] at (4.75,11.5) {$L_{k_1}$};
\node [font=\large] at (7.75,11.5) {$L_{k_2}$};
\node [font=\large] at (6.5,7.75) {$L_{(k_1 k_2)^{-1}}$};
\end{circuitikz}
&
\begin{circuitikz}
\tikzstyle{every node}=[font=\large]
\node at (6.25,10.5) [circ] {};
\node at (6.25,9.25) [circ] {};
\node at (6.25,10.5) [circ] {};
\draw [->, >=Stealth] (6.25,10.5) -- (5.5,11.25);
\draw [->, >=Stealth] (6.25,10.5) -- (7,11.25);
\draw [short] (5.5,11.25) -- (5,11.75);
\draw [short] (7,11.25) -- (7.5,11.75);
\node [font=\large] at (8.25,10.25) {$\sigma_{k_1 k_2} \times \mu(k_1,k_2)$};
\node [font=\large] at (7.5,9) {$\sigma_{(k_1k_2)^{-1}}$};
\draw [->, >=Stealth] (6.25,8) -- (6.25,8.75);
\draw [short] (6.25,8.75) -- (6.25,9.25);
\node [font=\large] at (5.25,11) {$L_{k_1}$};
\node [font=\large] at (7.25,11) {$L_{k_2}$};
\node [font=\large] at (6.5,7.75) {$L_{(k_1 k_2)^{-1}}$};
\end{circuitikz}
&
\begin{circuitikz}
\tikzstyle{every node}=[font=\large]
\node at (6.25,10.5) [circ] {};
\node at (6.25,10.5) [circ] {};
\draw [->, >=Stealth] (6.25,10.5) -- (5.5,11.25);
\draw [->, >=Stealth] (6.25,10.5) -- (7,11.25);
\draw [short] (5.5,11.25) -- (5,11.75);
\draw [short] (7,11.25) -- (7.5,11.75);
\node [font=\large] at (8.25,10.25) {$\sigma_{1} \times \mu(k_1,k_2)$};
\node [font=\large] at (5.25,11) {$L_{k_1}$};
\node [font=\large] at (7.25,11) {$L_{k_2}$};
\node [font=\large] at (6.5,7.75) {$L_{(k_1 k_2)^{-1}}$};
\draw [short] (6.25,9.25) -- (6.25,10.5);
\draw [->, >=Stealth] (6.25,8.25) -- (6.25,9.25);
\end{circuitikz}
    \end{tabular}
    \caption{The TPO's $\sigma_{k_i}$ inherit a product from a choice of Frobenius multiplication $\mu:H \otimes H \to H$ via the inclusion $L_{k_i} \subset H=\bigoplus_{\gamma \in \Gamma}L_{\gamma}$.}
    \label{fig:tpo-fusion}
\end{figure}

These two structures on the collection os TPO's, namely, the TPO product and the action by the TDL's, are compatible as in Figure~\ref{fig:tdl-tpo-module-map} (cf. \cite[Figure 13]{Robbins:2022wlr}). 
\begin{figure}[!h]
    \centering
  \begin{tabular}{cc}
\resizebox{0.4\textwidth}{!}{%
\begin{circuitikz}
\tikzstyle{every node}=[font=\large]
\node at (7.5,8.5) [circ, color={rgb,255:red,255; green,255; blue,255}] {};
\node at (5,8.5) [circ] {};
\node at (10,8.5) [circ] {};
\draw  (7.5,8.5) circle (3.75cm);
\draw [->, >=Stealth] (7.75,4.75) -- (7.25,4.75);
\draw [->, >=Stealth] (5,8.5) -- (5,10.5);
\draw [->, >=Stealth] (10,8.5) -- (10,10.5);
\draw [short] (5,10.5) -- (5,12.25);
\draw [short] (10,10.5) -- (10,12.25);
\node [font=\LARGE] at (5,8) {$\sigma_{k_1}$};
\node [font=\LARGE] at (5.75,9.75) {$L_{k_1}$};
\node [font=\LARGE] at (10,8) {$\sigma_{k_2}$};
\node [font=\LARGE] at (9.25,9.75) {$L_{k_2}$};
\node [font=\LARGE] at (7.5,4.25) {$L_{\gamma}$};
\node [font=\LARGE] at (4.75,12) {$L_{{\rm ad}_{\gamma}(k_1)}$};
\node [font=\LARGE] at (11.25,12) {$L_{{\rm ad}_{\gamma}(k_2)}$};
\end{circuitikz}
}$(a)$
&
\resizebox{0.4\textwidth}{!}{%
\begin{circuitikz}
\tikzstyle{every node}=[font=\LARGE]
\node at (7.5,8.5) [circ, color={rgb,255:red,255; green,255; blue,255}] {};
\node at (5,8.5) [circ] {};
\node at (10,8.5) [circ] {};
\draw [ color={rgb,255:red,255; green,255; blue,255} ] (7.5,8.5) circle (3.75cm);
\draw [ color={rgb,255:red,255; green,255; blue,255}, ->, >=Stealth] (7.75,4.75) -- (7.25,4.75);
\draw [->, >=Stealth] (5,8.5) -- (5,10.5);
\draw [->, >=Stealth] (10,8.5) -- (10,10.5);
\draw [short] (5,10.5) -- (5,12.25);
\draw [short] (10,10.5) -- (10,12.25);
\node [font=\LARGE] at (5,8) {$\sigma_{k_1}$};
\node [font=\LARGE] at (5.5,9) {$L_{k_1}$};
\node [font=\LARGE] at (10,8) {$\sigma_{k_2}$};
\node [font=\LARGE] at (10.5,9) {$L_{k_2}$};
\node [font=\LARGE, color={rgb,255:red,255; green,255; blue,255}] at (7.5,4.25) {$L_{\gamma}$};
\node [font=\LARGE] at (6.25,12) {$L_{{\rm ad}_{\gamma}(k_1)}$};
\node [font=\LARGE] at (11.25,12) {$L_{{\rm ad}_{\gamma}(k_2)}$};
\draw  (5,8.5) circle (1.25cm);
\draw  (10,8.5) circle (1.25cm);
\draw  (9.5,8.75) circle (0cm);
\draw  (10.75,5.25) circle (0cm);
\draw  (10.25,8.5) circle (0cm);
\draw  (9.25,9.5) circle (0cm);
\draw  (9.5,8.5) circle (0cm);
\draw  (7.75,8) circle (0cm);
\end{circuitikz}
} $(b)$ %
\\
\resizebox{0.4\textwidth}{!}{%
\begin{circuitikz}
\tikzstyle{every node}=[font=\large]
\node at (7.5,8.5) [circ] {};
\draw  (7.5,8.5) circle (3.75cm);
\draw [->, >=Stealth] (7.75,4.75) -- (7.25,4.75);
\draw  (9.5,8.75) circle (0cm);
\draw  (10.75,5.25) circle (0cm);
\draw  (14,13) circle (0cm);
\draw  (9.25,9.5) circle (0cm);
\draw  (9.5,8.5) circle (0cm);
\draw  (7.75,8) circle (0cm);
\node [font=\LARGE] at (7.5,4.25) {$L_{\gamma}$};
\draw [->, >=Stealth] (7.5,8.5) -- (7.5,12.25);
\draw [short] (7.5,12.25) -- (7.5,13.5);
\node [font=\LARGE] at (8.5,10.25) {$L_{k_1 k_2}$};
\node [font=\LARGE] at (7.75,8) {$\mu(k_1,k_2) \sigma_{k_1 k_2}$};
\node [font=\LARGE] at (9.75,13) {$L_{{\rm ad}_{\gamma}(k_1 k_2)}$};
\end{circuitikz}
} $(c)$ %
&
\resizebox{0.4\textwidth}{!}{
\begin{circuitikz}
\tikzstyle{every node}=[font=\large]
\node at (7.5,8.5) [circ, color={rgb,255:red,255; green,255; blue,255}] {};
\node at (5,8.5) [circ] {};
\node at (10,8.5) [circ] {};
\draw [ color={rgb,255:red,255; green,255; blue,255} ] (7.5,8.5) circle (3.75cm);
\draw [ color={rgb,255:red,255; green,255; blue,255}, ->, >=Stealth] (7.75,4.75) -- (7.25,4.75);
\draw [->, >=Stealth] (5,8.5) -- (5,10.5);
\draw [->, >=Stealth] (10,8.5) -- (10,10.5);
\draw [short] (5,10.5) -- (5,12.25);
\draw [short] (10,10.5) -- (10,12.25);
\node [font=\LARGE] at (5,8) {$\sigma_{{\rm ad}_{\gamma}(k_1)}$};
\node [font=\LARGE] at (10,8) {$\sigma_{{\rm ad}_{\gamma}(k_2)}$};
\node [font=\LARGE, color={rgb,255:red,255; green,255; blue,255}] at (7.5,4.25) {$L_{\gamma}$};
\node [font=\LARGE] at (6.25,12) {$L_{{\rm ad}_{\gamma}(k_1)}$};
\node [font=\LARGE] at (11.25,12) {$L_{{\rm ad}_{\gamma}(k_2)}$};
\draw  (9.5,8.75) circle (0cm);
\draw  (10.75,5.25) circle (0cm);
\draw  (10.25,8.5) circle (0cm);
\draw  (9.25,9.5) circle (0cm);
\draw  (9.5,8.5) circle (0cm);
\draw  (7.75,8) circle (0cm);
\end{circuitikz}
}$(d)$ 
\\ & \\ &  \\ 
\resizebox{0.4\textwidth}{!}{%
\begin{circuitikz}
\tikzstyle{every node}=[font=\LARGE]
\node at (7.5,8.5) [circ] {};
\draw  (9.5,8.75) circle (0cm);
\draw  (10.75,5.25) circle (0cm);
\draw  (14,13) circle (0cm);
\draw  (9.25,9.5) circle (0cm);
\draw  (9.5,8.5) circle (0cm);
\draw  (7.75,8) circle (0cm);
\draw [->, >=Stealth] (7.5,8.5) -- (7.5,12.25);
\draw [short] (7.5,12.25) -- (7.5,13.5);
\node [font=\LARGE] at (7.75,8) {${\rm ad}_{\gamma}\left(\mu(k_1,k_2) \sigma_{k_1 k_2}\right)$};
\node [font=\LARGE] at (9.75,13) {$L_{{\rm ad}_{\gamma}(k_1 k_2)}$};
\end{circuitikz}
}

$(e)$%
&
  \end{tabular}
    \caption{The TDL action on the TPO's distributes over the fusion of the TPO's, so that the compositions $(a)\to(c)\to(e)$ and $(a)\to (b)\to (d)\to (e)$ coincide.}
    \label{fig:tdl-tpo-module-map}
\end{figure}

Finally, the projector operators responsible for decomposition are obtained as the linear combinations of TPO's which are orthonormal under the TPO fusion product, and are invariant under the TDL action.

\subsection{Projection operators for Hopf algebra abelian extensions}\label{ssec:projection-operators}

We now construct the projector topological operators for our setting (\ref{eq:general-abeliansequence})
\begin{equation}
    {\rm Rep}(G) \xrightarrow{\imath} {\rm Rep}(H) \xrightarrow{F} {\rm Vec}(\Gamma)
 \end{equation}
of a trivially-acting category ${\rm Rep}(G)$, effectively-acting category ${\rm Vec}(\Gamma)$, and $H$ a Hopf algebra abelian extension (\ref{eq:abextension})
 \begin{equation}
     \C^{\Gamma} \to H \to \C G.
 \end{equation}

To do this, we first observe that the structures coming from the physical arguments summarized above can be described in more algebraic terms as follows:
\begin{enumerate}
    \item Each simple object $\rho \in {\rm Rep}(G)$ of the trivially-acting category defines a ${\rm dim}(\rho)$-dimensional vector space $\Sigma_{\rho}$ of TPO's via the junction hom-space
    \begin{equation}
       \Sigma_{\rho}:= {\rm Hom}\left(\mathbbm{1}, F\left( \imath \left( \rho \right) \right) \right).
    \end{equation}
    This is Figure~\ref{diagram:tpo-definition}, except that now we have ${\rm dim}(\rho)$ linearly-independent TPO's.
    \item Given a choice of symmetric special Frobenius algebra structure $(\mu_F,\Delta_F)$ on the regular object $H^*$ of ${\rm Rep}(H)$, the (embedding of the) regular representation $R$ in ${\rm Rep}(G)$ is itself an algebra $(R,\mu_F\vert_R: R \otimes R \to R)$. This, in turn, endows the ${\rm dim}(R) = \vert G \vert$-dimensional vector space $\Sigma_R$ of TPO's
     \begin{equation}
       \Sigma_R:= {\rm Hom}\left(\mathbbm{1}, F\left( \imath \left( R \right) \right) \right)
    \end{equation}
    with the same algebra structure. This is the direct generalization of Figure~\ref{fig:tpo-fusion}, schematically presented in Figure~\ref{fig:tpo-fusion-regular}, where $\{\sigma_R\}$ stands for the ${\rm dim}(R) = \vert G \vert$-dimensional vector space $\Sigma_R$ of TPO's. Of course, for computations, as we already saw in Section~\ref{sec:pfuncdecomposition}, one needs to expand on some basis, this we will do below.
    \begin{figure}[]
    \centering
    \begin{tabular}{ccc}
   \begin{circuitikz}
\tikzstyle{every node}=[font=\large]
\node at (5,10.5) [circ] {};
\node at (6.25,9.25) [circ] {};
\node at (7.5,10.5) [circ] {};
\draw [->, >=Stealth] (5,10.5) -- (4.25,11.25);
\draw [->, >=Stealth] (7.5,10.5) -- (8.25,11.25);
\draw [short] (4.25,11.25) -- (3.75,11.75);
\draw [short] (8.25,11.25) -- (8.75,11.75);
\node [font=\large] at (5,10.25) {$\{\sigma_{R}\}$};
\node [font=\large] at (7.5,10.25) {$\{\sigma_{R}\}$};
\node [font=\large] at (7.5,9) {$\{\sigma_{R}\}$};
\draw [->, >=Stealth] (6.25,8) -- (6.25,8.75);
\draw [short] (6.25,8.75) -- (6.25,9.25);
\node [font=\large] at (4.75,11.5) {$R$};
\node [font=\large] at (7.75,11.5) {$R$};
\node [font=\large] at (6.5,7.75) {$R$};
\end{circuitikz}
&
\begin{circuitikz}
\tikzstyle{every node}=[font=\large]
\node at (6.25,10.5) [circ] {};
\node at (6.25,9.25) [circ] {};
\node at (6.25,10.5) [circ] {};
\draw [->, >=Stealth] (6.25,10.5) -- (5.5,11.25);
\draw [->, >=Stealth] (6.25,10.5) -- (7,11.25);
\draw [short] (5.5,11.25) -- (5,11.75);
\draw [short] (7,11.25) -- (7.5,11.75);
\node [font=\large] at (8.25,10.25) {$\left(\mu_F\vert_R\right)\{\sigma_{R}\}$};
\node [font=\large] at (7.5,9) {$\{\sigma_{R}\}$};
\draw [->, >=Stealth] (6.25,8) -- (6.25,8.75);
\draw [short] (6.25,8.75) -- (6.25,9.25);
\node [font=\large] at (5.25,11) {$R$};
\node [font=\large] at (7.25,11) {$R$};
\node [font=\large] at (6.5,7.75) {$R$};
\end{circuitikz}
&
\begin{circuitikz}
\tikzstyle{every node}=[font=\large]
\node at (6.25,10.5) [circ] {};
\node at (6.25,10.5) [circ] {};
\draw [->, >=Stealth] (6.25,10.5) -- (5.5,11.25);
\draw [->, >=Stealth] (6.25,10.5) -- (7,11.25);
\draw [short] (5.5,11.25) -- (5,11.75);
\draw [short] (7,11.25) -- (7.5,11.75);
\node [font=\large] at (8.25,10.25) {$\left(\mu_F\vert_R\right)\{\sigma_{R}\}$};
\node [font=\large] at (5.25,11) {$R$};
\node [font=\large] at (7.25,11) {$R$};
\node [font=\large] at (6.5,7.75) {$R$};
\draw [short] (6.25,9.25) -- (6.25,10.5);
\draw [->, >=Stealth] (6.25,8.25) -- (6.25,9.25);
\end{circuitikz}
    \end{tabular}
    \caption{The ${\rm dim}(R) = \vert G \vert$-dimensional vector space $\Sigma_R$ of TPO's $\{\sigma_R\}$ inherit a product from a choice of Frobenius multiplication $\mu:H \otimes H \to H$ on the regular object of ${\rm Rep}(H)$ via the inclusion $R \subset H$.}
    \label{fig:tpo-fusion-regular}
\end{figure}
\item The TDL's in ${\rm Rep}(H)$ act on the space $\Sigma_R$ of TPO's (Figures \ref{fig:tdl-action}, \ref{fig:effective-tdl-action}) or, phrased in another way, the TPO's carry a $H^*$-module structure, for $H^*$ the Hopf algebra dual to $H$. 
\item Finally, Figure~\ref{fig:tdl-tpo-module-map} says that the product of $\Sigma_R$ is a $H^*$-module map. 
\end{enumerate}

All in all, we see that a choice of a symmetric special Frobenius algebra that gauges ${\rm Rep}(H)$ in particular ought to define a $\vert G \vert$-dimensional algebra object of TPO's $\Sigma_R$ in ${\rm Rep}(H^*)$, that is, a  $\vert G \vert$-dimensional vector space that is a $H^*$-module and is equipped with a product $\Sigma_R \otimes \Sigma_R \to \Sigma_R$ which is furthermore a map of $H^*$-modules.

This is another instance where it pays off to work explicitly with the Hopf algebras themselves. Indeed, it suffices to dualize the sequence (\ref{eq:abextension}), yielding another extension of Hopf algebras
\begin{equation}
    \C^G \to H^* \to \C\Gamma.
\end{equation}
By definition, the algebra product on the dual Hopf algebra $H^*$ is the dual of the coproduct of $H$, which is the Frobenius multiplication (\ref{eq:frobeniusmultisadual}) used to gauge ${\rm Rep}(H)$. The algebra $\C^G$, which is the regular representation $R$ in the trivially-acting category ${\rm Rep}(G)$, is therefore identified with the algebra $\Sigma_R$ of TPO's. Now, $H^*$ acts on itself via the adjoint action, and by exactness of the sequence, $\C^G \hookrightarrow H^*$ is a normal subalgebra of $H^*$, meaning it is a $H^*$-module\footnote{In \cite{Robbins:2022wlr}, \textit{projective} adjoint actions were considered. In some cases, but not always, the projective (there also referred to as ``mixed anomaly'') phases were observed to coincide with turning on discrete torsion, which amounts to changing the Frobenius algebra one uses to gauge the symmetry. Here, by contrast, the Frobenius algebra is fixed to be the one constructed from the Hopf algebra itself, as we are chiefly concerned with obtaining the decomposition projector operators. It would however be interesting to delve deeper into the relation of projective adjoint actions of Hopf algebras and mixed anomalies. \label{footnote:mixed-anomaly-phases}} whose product is a map of $H^*$-modules. 

We note in passing that this formulation recovers the algebra (with trivial discrete torsion on $\Gamma$) and module structures on the vector space of TPO's described in Section~\ref{ssec:tpo-invertible}, by working with the exact sequence of Hopf group algebras induced by (\ref{eq:tdl-gp-extension-review})
\begin{equation}
    \C K \hookrightarrow \C \Gamma \to \C G,
\end{equation}
by setting $H = \C^{\Gamma}$, since the symmetry category is ${\rm Vec}(\Gamma) \cong {\rm Rep}(\C^{\Gamma})$ and $(\C^{\Gamma})^* = \C\Gamma$. In that case, the TPO's $\{\sigma_k\}_{k \in K}$ correspond to the usual basis elements $\{k \in K\}\subset \C K$ spanning the algebra.

The only step left to construct the projection operators is to identify the orthonormal elements of $\C^G$ which are invariant under the $H^*$-action. Since the definition of the adjoint action requires knowing the Hopf algebra structure $(u_d:\C\to H^*,\mu_d: H^* \otimes H^* \to H^*,\epsilon_d: H^* \to \C, \Delta_d: H^* \to H^* \otimes H^*, S_d: H^* \to H^*)$ on $H^*$, which we construct by taking the dual maps of (\ref{eq:abextunit})-(\ref{eq:abantipode}), let us list it explicitly in the basis $\{ g \# v_x\}_{g\in \Gamma, x\in G}$ dual to that of $H$. The algebra structure and part of the coalgebra structure had already been computed above (cf. Eq's.~(\ref{eq:frobeniusmultisadual})-(\ref{eq:frobenius-unit-dual}), (\ref{eq:module-decompo-coprod})).
\begin{equation}
    u_d:=\epsilon^*: 1 \mapsto \sum_{x\in G} 1 \# v_x,
\end{equation}
\begin{equation}
    \mu_d:= \Delta^* : \left(g\# v_x\right)\otimes \left(h\# v_y\right) \mapsto \tau_y(g,h) gh\# \delta_{x,h\triangleright y}v_y,\label{eq:dualhopfproduct}
\end{equation}
\begin{equation}
    \epsilon_d:= u^*: (g \# v_x) \mapsto \delta_{1,x}m
\end{equation}
\begin{equation}
    \Delta_d:=\mu^*: g \# v_x \mapsto \sum_{y\in G} \sigma_g(y, y^{-1}x) \left(\left(g \# v_y \right) \otimes \left(\left( g \triangleleft y\right) \# v_{y^{-1}x} \right) \right),
\end{equation}
\begin{equation}
    S_d:= S^*: g \# v_x \mapsto \left(\sigma_g(x,x^{-1}) \, \tau_{(g \triangleright x)^{-1}}\left(g \triangleleft x, \left(g \triangleleft x \right)^{-1} \right) \right)^{-1} \left(\left( g \triangleleft x\right)^{-1} \# v_{(g \triangleright x)^{-1}} \right).
\end{equation}

Now, the adjoint action\footnote{Here, we consider the right adjoint action for concreteness. One can also work with the left adjoint action, where the antipode is applied to $a_{(2)}$ instead of to $a_{(1)}$.} takes the general form
\begin{equation}\label{eq:dual-adjoint-action-general}
    {\rm ad}^r_{a}(b) = S_d(a_{(1)}) b a_{(2)},
\end{equation}
for $a,b \in H^*$ and where we used Sweedler's notation (see Appendix~\ref{app:sweedler}) to write the comultiplication $a_{(1)}\otimes a_{(2)} = \Delta_d(a) \in H^* \otimes H^*$, with the sum left implicit.

The algebra inclusion $\imath : \C^G \hookrightarrow H^*$ is, in the dual basis,
\begin{equation}
    \imath: v_x \mapsto 1 \# v_x.
\end{equation}

Notice that, according to the product (\ref{eq:dualhopfproduct}) on $H^*$, the basis elements $\{1 \# v_x \}_{x \in G}$ are already orthonormal, so one only needs to compute the adjoint-invariant combinations. On a given basis element $1 \# v_z$, the adjoint action (\ref{eq:dual-adjoint-action-general}) can be computed as
\begin{eqnarray}
    {\rm ad}^r_{g \# v_x}(1 \# v_z) &=& \sum_{y\in G} \Big(\sigma_g(y,y^{-1}x) \left(\sigma_g(y,y^{-1}) \tau_{(g \triangleright y)^{-1}}(g \triangleleft y, (g \triangleleft y)^{-1}) \right)^{-1} \delta_{(g \triangleright y)^{-1},z}\nonumber
    \\
    && \tau_z( (g \triangleleft y)^{-1},1) \delta_{z,(g \triangleleft y)\triangleright (y^{-1}x)} \tau_{y^{-1}x}((g \triangleleft y)^{-1},g \triangleleft y)\Big) \left( 1 \# v_{y^{-1}x}\right)
    \\
    &=& \delta_{1,x} \left( 1 \# v_{\left(g^{-1} \triangleleft z^{-1} \right)\triangleright z}\right).
\end{eqnarray}
It is noteworthy that the cocycles $\tau,\sigma$ do not contribute to the adjoint action.

Thus, we observe that the ad-invariant linear combinations of TPO's are given by
\begin{equation}\label{eq:projectors-formula}
    \Pi_{[x]}:= \sum_{z\in [x]} 1\# v_z,
\end{equation}
for $[x]\in G/\Gamma$. It is straightforward to see that
\begin{equation}
    \Pi_{[x]} \Pi_{[y]} = \delta_{[x],[y]} \Pi_{[x]}
\end{equation}
\begin{equation}
    \sum_{[x]\in G/\Gamma} \Pi_{[x]} = \sum_{z\in G} 1\# v_z = u_d(1) = \mathbbm{1},
\end{equation}
so that the collection $\{ \Pi_{[x]} \}_{x \in G /\Gamma}$ of TPO's are the projectors enforcing the decomposition into the family of $G/\Gamma$ universes, as predicted by the decomposition conjecture (\ref{eq:general-abconjecture}).

\section{Examples}\label{sec:examples}

\subsection{Group-like case: abelian kernel}\label{ssec:gplikecase}

As a first example, we specialize to the extension of a finite group $\Gamma$ by a finite abelian group $G$. Group-theoretically, this is an exact sequence of groups of the form
\begin{equation*}
    1\to G\to \widetilde{G} \to \Gamma \to 1,
\end{equation*}
specified by a $\Gamma$-action on $G$ and a 2-cocycle of $\Gamma$ valued in $G$. 

Our prediction (\ref{eq:general-abconjecture}) in this case reduces to that of \cite{Hellerman:2006zs} for a trivially-acting normal abelian group $G$:
\begin{equation}\label{eq:gplikeprediction}
   [{\cal T}/ \widetilde{G}] = \bigoplus_{[\rho]\in G/\Gamma} [{\cal T}/H_{\rho}]_{\tau_{\rho}}.
\end{equation}
using the identification $\text{Irrep}(G)= \text{Hom}(G,U(1))= G$ \textit{as $\Gamma$-sets.}

The group exact sequence information is captured by the dual group algebra exact sequence
\begin{equation}
    \C^{\Gamma}\to \C^{\widetilde{G}} \to \C^G,
\end{equation}
which turns into an abelian extension (\ref{eq:abextension}) of Hopf algebras after \textit{choosing an isomorphism} $p: \C^G \xrightarrow{\sim} \C \hat{G}$ for $\hat{G}=\text{Hom}(G,U(1))$ the Pontryagin dual of $G$. In this case, both the action by permutation $\Gamma\times \hat{G} \to \Gamma$ as well as the 2-cocycle $\sigma: \hat{G} \times \hat{G} \to (\C^{\times})^{\Gamma}$ are trivial.

Thus, we can use our general result to describe abelian extensions of groups
\begin{equation}
    \text{Rep}(\hat{G})\to \text{Vec}(\widetilde{G}) \xrightarrow{F} \text{Vec}(\Gamma).
\end{equation}
The $2$-torus partition function for gauging $\text{Vec}(\widetilde{G})$ with $\text{Rep}(\hat{G})$ trivially acting according to Eq.~(\ref{eq:2toruspartitionfunction}) is
\begin{equation}
    Z_{\widetilde{G}} = \sum_{[\rho]\in \hat{G}/\Gamma} Z_{K_{\rho},\rho\circ\tau},
\end{equation}
for $Z_{K_{\rho},\rho\circ\tau}$ a $K_{\rho}\leq \Gamma$ orbifold with discrete torsion $\rho\circ\tau \in Z^2(K_{\rho},U(1))$, which recovers the decomposition conjecture of groups (\ref{eq:gplikeprediction}). The projectors (\ref{eq:projectors-formula}) in this case are
\begin{equation}
    \Pi_{[\rho]} = \sum_{s\in [\rho]} 1 \# v_s,
\end{equation}
where $v_s$ is identified with the idempotent corresponding to the irrep $s \in \hat{G}$.

\subsection{Group representations}

We now specialize to a sequence of group representation categories
\begin{equation}
    \text{Rep}(G)\to\text{Rep}(\widetilde{G})\to \text{Rep}(\Gamma),
\end{equation}
for $\Gamma$ a finite abelian group. A monoidal equivalence $\text{Rep}(\Gamma)\cong \text{Vec}(\Gamma)$ thus gives a sequence 
\begin{equation}
    \text{Rep}(G)\to\text{Rep}(\widetilde{G})\to \text{Vec}(\Gamma),
\end{equation}
for which the decomposition conjecture (\ref{eq:general-abconjecture}) predicts
\begin{equation}\label{eq:repcaseprediction}
    [{\cal T}/\text{Rep}(H)] = \bigoplus_{i\in G} [{\cal T}/\Gamma].
\end{equation}

In more technical terms, we start with an exact sequence of groups
\begin{equation*}
    1\to \Gamma \to \widetilde{G}\to G\to 1,
\end{equation*}
determined by a $G$-action on $\Gamma$, and a 2-cocycle on $G$ valued in $\Gamma$. If $\Gamma$ is an abelian group, then we can \textit{choose} a Hopf isomorphism to obtain an abelian extension of Hopf algebras
\begin{equation}
    \C\Gamma\xrightarrow{\sim}\C^{\Gamma}\to \C\widetilde{G}\to \C G.
\end{equation}
In this case, the $\Gamma$-action by permutation on $G$ and the 2-cocycle on $\Gamma$ valued in $G$ are trivial, and this gives an exact sequence of fusion categories
\begin{equation}
    \text{Rep}(G)\to \text{Rep}(\widetilde{G})\to \text{Rep}(\Gamma)\cong \text{Vec}(\Gamma).
\end{equation}
Intuitively, this scenario is dual to the one studied in Section~\ref{ssec:gplikecase}.

The partition function splits then as
\begin{equation}
    Z_{(\C\widetilde{G})^*}= \sum_{x\in G} Z_{\hat{\Gamma}} = |G|\, Z_{\Gamma},
\end{equation}
where $Z_{(\C\widetilde{G})^*}$ means gauging by the Frobenius algebra constructed from the fiber functor corresponding to the Hopf algebra $\C\widetilde{G}$.

This agrees with the decomposition conjecture (\ref{eq:repcaseprediction}) and the various results derived in \cite[Section 5]{Perez-Lona:2023djo}, \cite[Section 6]{Perez-Lona:2024sds}. This is further supported by computing the projectors (\ref{eq:projectors-formula}). Since the $\Gamma$-action on $G$ is trivial, then each equivalence class $[g]\in G/\Gamma$ has exactly one element (namely, $g\in [g]$), so that the projectors are simply $\Pi_{g\in G}$ with $\Pi_g \Pi_{h} = \delta_{g,h}\Pi_g$.

\subsection{Extensions with $G=\Z_2$}
We now consider the special case where $G=\Z_2$ but, in contrast with Section~\ref{ssec:gplikecase}, allow for the most general extension information $(\triangleleft, \triangleright,\sigma,\tau)$. In this case, the abelian extension $H$ in
\begin{equation}
    \C^{\Gamma}\to H\to \C\Z_2
\end{equation}
will not necessarily be commutative nor cocommutative. This describes the case of 
\begin{equation}
    \text{Rep}(\Z_2)\to \text{Rep}(H)\to \text{Vec}(\Gamma)
\end{equation}
a $\text{Rep}(\Z_2)$-trivially acting subcategory and a $\text{Vec}(\Gamma)$ the remaining symmetry category, so that we can see discrete torsion emerge in the decomposition.

The decomposition conjecture (\ref{eq:general-abconjecture}) predicts that
\begin{equation}\label{eq:z2hopfexampleconj}
    [{\cal T}/\text{Rep}(H)] = [{\cal T}/\Gamma] \oplus [{\cal T}/\Gamma]_{\omega},
\end{equation}
for $\omega\in Z^2(\Gamma,U(1))$ a $\Gamma$-discrete torsion determined by the sequence.
This clear-cut decomposition is due to the specific properties of $\Z_2$ Hopf extensions, as we describe in further detail now. 

\subsubsection{Extension data}

The properties of $\Z_2:=\langle x\vert x^2=1\rangle$ allows us to considerably simplify the extension information. First, note that the identity (\ref{eq:trianglerightidentity}) implies $g\triangleright 1=1$ for all $g\in \Gamma$, so that $g\triangleright x=x$. In turn, the identity (\ref{eq:triangleleftifentity}) implies that $\Z_2$ acts on $\Gamma$ via \textit{group} homomorphisms.

Now, the requirement (\ref{eq:2cocyclesevaluatedatone}) on $\tau$ implies that only $\tau_x(g,h)$ can be nontrivial. Combining with the 2-cocycle identity (\ref{eq:tau2cocycle}) shows that the $\tau$-cocycle in this case simply encodes a choice of discrete torsion $\tau_x\in Z^2(\Gamma,U(1))$ on $\Gamma$. Furthermore, applying the normalization condition (\ref{eq:sigmanormalization}) to $\sigma$ says that only the function $\sigma(x,x)\in (\C^{\times})^{\Gamma}$ can be nontrivial, and so this cocycle is simply a function $\sigma: \Gamma\to \C^{\times}$.

Therefore, the extension information in this case reduces to \cite{ZL21}
\begin{enumerate}
    \item A group automorphism $- \triangleleft x: \Gamma\xrightarrow{\sim}\Gamma$,
    \item A choice of discrete torsion $\tau_x\in Z^2(\Gamma,U(1))$ on $\Gamma$,
    \item A function $\sigma: \Gamma\to \C^{\times}$, which together with $\tau$ satisfies the condition
    \begin{equation}
        \sigma(st)=\sigma(s)\sigma(t)\tau_x(s,t)\tau_x(s\triangleleft x,t\triangleleft x).
    \end{equation}
    \end{enumerate}

With this information, it is straightforward to compute the partition function. This takes the form
\begin{equation}
    Z_{H} =  \left(\frac{1}{\vert\Gamma \vert}\sum_{g,h\in \Gamma}\delta_{gh,hg} \, Z_{g,h} \right) + \left(\frac{1}{\vert\Gamma \vert}\sum_{g,h\in\Gamma} \delta_{gh,hg} \, \frac{\tau_x(g,h)}{\tau_x(h,g)} \, Z_{g,h}\right) = Z_{\Gamma} + Z_{\Gamma,\tau_x},
\end{equation}
which agrees with the decomposition prediction (\ref{eq:z2hopfexampleconj}) that the $\text{Rep}(H)$-orbifold with a trivially-acting $\text{Rep}(\Z_2)$ subsymmetry splits as two $\Gamma$-orbifolds, one without discrete torsion, and one with the choice of discrete torsion $\tau_x\in Z^2(\Gamma,U(1))$.

Notice that a consequence of the triviality of the $\Gamma$-action is that we get exactly $\vert G \vert = \vert \Z_2 \vert = 2$ projectors according to the general derivation in (\ref{eq:projectors-formula}), namely
\begin{equation}\label{eq:kac-pal-proj}
    \{\Pi_x\}_{x\in \Z_2} \quad {\rm s.t.} \quad \Pi_x\Pi_y = \delta_{x,y} \Pi_x.
\end{equation}

Let us now connect to a specific, more familiar example.

\subsubsection{Generalized Kac-Paljutkin algebra with trivially acting $\Z_2$}
The Kac-Paljutkin Hopf algebra $H_8$ is the lowest-dimensional semisimple Hopf algebra that is neither commutative nor cocommutative \cite{KP66}. It is the $n=2$ member of a family of Hopf abelian extensions $H_{2n^2}$  \cite{Pan17} of dimension $2n^2$ of the form
\begin{equation}\label{eq:h8abelianextension} 
    \C^{\Z_n\times\Z_n} \to H_{2n^2} \to \C\Z_2.
\end{equation}
Denoting the generator of $\Z_2$ as $x$, those of $\Z_2\times\Z_2$ as $\{g,h\}$, and $w$ an $n$th root of unity in $\C$, the extension data takes the form \cite{ZL21}
\begin{enumerate}
    \item a \textit{group} homomorphism $a\triangleleft x = b$, $b\triangleleft x = a$ ,
    \item a function $\sigma(a^i b^j) = w^{ij}$,
    \item a 2-cocycle 
    \begin{equation}
        \tau_x(a^i b^j, a^kb^l)= w^{jk}. \label{eq:znzndt}
    \end{equation}
\end{enumerate}

With this information at hand, we can compute the partition function for a gauged $\text{Rep}(H_{2n^2})$ symmetry with its Frobenius algebra determined by the fiber functor corresponding to $H_{2n^2}$, and a trivially-acting $\Z_2$ subsymmetry. The partition function takes the form
\begin{gather}
    Z_{H_{2n^2}} = \left(\frac{1}{\vert\Z_n\times\Z_n\vert}\sum_{0\leq i,j,k,l<n} Z_{a^ib^j,a^kb^l} \right) + \left(\frac{1}{\vert\Z_n\times\Z_n\vert}\sum_{0\leq i,j,k,l<n} w^{jk-il} \, Z_{a^ib^j,a^kb^l}\right), \\ = Z_{\Z_n\times\Z_n} + Z_{\Z_n\times \Z_n,\tau_x},
\end{gather}
which agrees with the decomposition prediction (\ref{eq:z2hopfexampleconj})
\begin{equation}
    [{\cal T}/\text{Rep}(H_{2n^2})] = [{\cal T}/\Z_n\times\Z_n]\oplus[{\cal T}/\Z_n\times \Z_n]_{\tau_x} 
\end{equation}
that the $\text{Rep}(H_{2n^2})$-gauged theory splits as a $(\Z_n\times\Z_n)$-orbifold without discrete torsion, and a $(\Z_n\times\Z_n)$-orbifold with discrete torsion given by $\tau_x:(\Z_n\times \Z_n)\times (\Z_n\times \Z_n)\to U(1)$ given by (\ref{eq:znzndt}).

At the level of projectors (\ref{eq:kac-pal-proj}), one can see that the projector $\Pi_1$ is associated with the universe without discrete torsion, whereas the projector $\Pi_x$ corresponds to that with discrete torsion $\tau_x$.

In particular, the case of the Kac-Paljutkin algebra $H_8$ then splits as
\begin{equation}
    [{\cal T}/\text{Rep}(H_{8})] = [{\cal T}/\Z_2\times\Z_2]\oplus[{\cal T}/\Z_2\times \Z_2]_{\tau_x},
\end{equation}
for $\tau_x$ a nontrivial choice of discrete torsion on $\Z_2\times\Z_2$.

\section{Extension to more general Hopf extensions}\label{sec:generalextensions}

\subsection{Decomposition conjecture}

As mentioned in Section~\ref{ssec:hopfexact} (cf. Appendix~\ref{sapp:hopfexact}), one can consider exact sequences of Hopf algebras
\begin{equation}
    H' \xrightarrow{i} H \xrightarrow{\pi} H'', 
\end{equation}
where neither $H'$ nor $H''$ is a (dual) group algebra. This is the mathematical setting describing a theory $\cal T$ with a $\text{Rep}(H)$ symmetry with a trivially-acting subsymmetry $\text{Rep}(H'')$ and a remaining symmetry $\text{Rep}(H')$
\begin{equation}
    \text{Rep}(H'')\to\text{Rep}(H)\to \text{Rep}(H'),
\end{equation}
where potentially all three fusion categories have non-invertible simple objects (but all admit SPT phases).

As shown explicitly in Section~\ref{ssec:gplikecase}, the decomposition conjecture for Hopf abelian extensions subsumes the case of a finite group symmetry with a trivially-acting normal \textit{abelian} group. However, the finite group decomposition conjecture \cite{Hellerman:2006zs} deals more generally with \textit{nonabelian} trivially-acting normal subgroups. There, the decomposition conjecture for the scenario
\begin{equation}
    \text{Vec}(N)\to \text{Vec}(\widetilde{G})\to \text{Vec}(\Gamma),
\end{equation}
(or the Hopf exact sequence
\begin{equation}\label{eq:dualgroupextension}
    \C^{\Gamma} \to \C^{\widetilde{G}}\to \C^{N},
\end{equation}
if we are interested in working with representation categories,) states that
\begin{equation}
    [{\cal T}/\widetilde{G}] = \bigoplus_{[\rho]\in \text{Irrep}(N)/G} [{\cal T}/K_{\rho}]_{\omega_{\rho}},
\end{equation}
where $\text{Irrep}(N)/\Gamma$ is the orbit \textit{set} of the irreducible representations of $N$, not necessarily all one-dimensional, by the action of $\Gamma$, $K_{\rho}$ is the subgroup of $\Gamma$ that acts trivially on the orbit of $\rho$, and $\omega_{\rho}$ is some discrete torsion. In particular, $\text{Irrep}(N)$ is only taken \textit{as a }$\Gamma$-\textit{set}, no reference to its ring structure is made.

On the other hand, the partition function computation for a Hopf abelian extension
\begin{equation}\label{eq:case1}
    \C^{\Gamma}\to H\to \C G,
\end{equation}
where in particular $H$ is not necessarily commutative anymore (as opposed to (\ref{eq:dualgroupextension})), with associated exact sequence
\begin{equation}\label{eq:case2}
    \text{Rep}(G)\to\text{Rep}(H)\to \text{Vec}(\Gamma),
\end{equation}
showed the interesting result that, at least for the partition function, only the coalgebra extension information plays a role in the decomposition. In this case, we have a sum
\begin{equation}
    [{\cal T}/H] = \bigoplus_{[x]\in G/\Gamma} [{\cal T}/K_x]_{\tau_x},
\end{equation}
where now the sum runs over the orbits of $G$ \textit{as a $\Gamma$-set}, the orbifolds are by some subgroup $K_x$ of $\Gamma$ that leaves the orbit of $x$ fixed, and $\tau_x$ is again some discrete torsion.

Combining these observations, \textit{with the proviso that the algebra information of} $H$ \textit{remains irrelevant} (as argued in Section~\ref{sec:general}), suggests the following decomposition conjecture: 

\begin{conj}
{ \rm
Given an exact sequence of Hopf algebras
\begin{equation}\label{eq:generalextension}
    H'\to H\to H'',
\end{equation}
with associated fusion categories
\begin{equation}
    \text{Rep}(H'')\to \text{Rep}(H)\to \text{Rep}(H'),
\end{equation}
describing a theory $\cal T$ with a $\text{Rep}(H)$-symmetry, a trivially-acting $\text{Rep}(H'')$-subsymmetry, and a remaining $\text{Rep}(H')$ symmetry, gauging $\text{Rep}(H)$ using the Frobenius algebra constructured from the fiber functor that reconstructs $H$, results in the decomposition
\begin{equation}
    [{\cal T}/\text{Rep}(H) ]= \bigoplus_{i\in \mathcal{O}} \, [{\cal T}/(K_i,\mu_i,\Delta_i)]
\end{equation}
consisting of a sum of orbifolds parameterized by the set of orbits $\mathcal{O}$ of the irreducible representations $\mathcal{I}((H'')^*)$ of the \textit{dual} Hopf algebra $(H'')^*$ by an action of the fusion category $\text{Rep}(H')$, such that for each orbit $i\in \mathcal{O}$, the associated orbifold is by the Hopf subalgebra $K_i\subset (H')^*$ which acts trivially on the orbit $i$, and the discrete torsion choice\footnote{In the sense of \cite{Perez-Lona:2024yih}, a (Morita class of) symmetric special Frobenius algebra structure on $K_i$.} $(\mu_i,\Delta_i)$ on $K_i$ is determined by the orbit.
}
\end{conj}

In the first case (\ref{eq:dualgroupextension}), $H''= \C^N$, so the irreducible representations of $(H'')^*=\C N$ are the irreps of $N$, and the action of $\text{Rep}(H') = \text{Vec}(\Gamma)$ is the (linear extension of) precomposition of an irrep by the $\Gamma$-action on $N$. In the second case (\ref{eq:case1}), the irreps of $(H'')^*=\C^G$ are simply the homogeneous $G$-graded one-dimensional vector spaces, and the action of $\text{Rep}(H')=\text{Vec}(\Gamma)$ is given by the action by permutation.

The two main questions therefore become how to describe this $\text{Rep}(H)$-action on the irreps of $(H'')^*$, and where the discrete torsion choice comes from. To address both questions, we will construct an action of $\text{Comod}((H')^*) = \text{Rep}(H')$ on $\text{Rep}((H'')^*)$ by \textit{linear} endofunctors. Requiring only linear endofunctors is the equivalent of requiring an action of $G$ on $\text{Irrep}(N)$, or $\Gamma$ on $G$, regarded as a \textit{set}. While this action will come with additional structure (see e.g. \cite[Section 8.2]{Nat20}) resembling a monoidal structure on each linear endofunctor, this data is determined by the algebra extension information, which was observed to be irrelevant for abelian extensions, and argued to remain so at the end of Section~\ref{ssec:frobeniusalgebra}. We leave an explicit verification of this in the spirit of Section \ref{sec:pfuncdecomposition} for future work.

In the next sections, we construct an explicit action functor, define the ``orbits'' of irreducible representations, and describe how the discrete torsion choice is obtained at this level of generality.

\subsection{Action tensor functor}

In this section, we describe a tensor functor
\begin{equation}
    \rho: \text{Comod}((H')^*)\to \text{End}(\text{Rep}((H'')^*))
\end{equation}
where $\text{End}(\text{Rep}((H'')^*))$, the linear category of linear endomorphisms of  $\text{Rep}((H'')^*)$, is regarded as a (strict) monoidal category with monoidal product given by composition.

As mentioned in Section~\ref{ssec:hopfexact}, the coalgebra constituents of the extension data of a more general Hopf exact sequence are a co-action of $H'$ on $H''$, and a 2-cocycle of $H'$ valued in $H''$. One can always dualize the exact sequence (\ref{eq:generalextension}) to obtain another exact sequence of finite-dimensional semisimple Hopf algebras
\begin{equation}
    (H'')^*\to H^* \to (H')^*.
\end{equation}
The extension data defining this Hopf extension is simply the dual\footnote{Generically, the dual extension data does not match the original extension data, but remarkably in some contexts it does (see e.g. \cite{AN99}).} to the extension of (\ref{eq:generalextension}), so that in particular we get a 2-cocycle $\tau: (H')^*\otimes (H')^*\to (H'')^*$ and a weak action $\triangleright: (H')^*\otimes (H'')^*\to (H'')^*$ satisfying the following conditions (cf. Section~\ref{ssec:hopfexact}):
\begin{enumerate}
    \item normalized 2-cocycle condition
    \begin{gather} \label{eq:2cocyclenormalization}
    \tau (1,h)= \tau(h,1)= \epsilon(h)\,1
    \\
    (h_{(1)}\triangleright \tau(l_{(1)},m_{(1)}) \tau(h_{(2)},l_{(2)}m_{(2)}) = \tau(h_{(1)},l_{(1)})\tau(h_{(2)}l_{(2)},m), \label{eq:2cocycletaucondition}
\end{gather}
\item weak action condition
\begin{gather}
    h\triangleright (ab) = (h_{(1)}\triangleright a)(h_{(2)}\triangleright b),
    \\
    h\triangleright 1 = \epsilon(h) \, 1,
    \\
    1\triangleright a = a.
\end{gather}
\item twisted module condition

\begin{equation}\label{eq:twistedmodulecondition}
    (h_{(1)}\triangleright (l_{(1)}\triangleright a))\tau(h_{(2)},l_{(2)}) = \tau(h_{(1)},l_{(1)})(h_{(2)}l_{(2)}\triangleright a).
\end{equation}
\end{enumerate}

\subsubsection{Underlying action functor}

Let $(V,c:V\to (H')^*\otimes V)$  be a $(H')^*$-comodule, and $(W,r:(H'')^*\otimes W\to W)$ a $(H'')^*$-representation. We can endow the vector space $V\otimes W$ with the following linear map
\begin{equation}
    r_{(V,c)}: (H'')^*\otimes (V\otimes W)\to V\otimes W,
\end{equation}
\begin{gather}
    r_{(V,c)}= \big((\text{id}_V\otimes r)(b_{(H'')^*), V}\otimes \text{id}_W\big)\circ \big( (\triangleright(S_{(H')^*}\otimes \text{id}_{(H'')^*})(b_{(H'')^*, (H')^*}))\otimes \text{id}_{V\otimes W})\big) \nonumber \\ \circ \big( \text{id}_{(H'')^*}\otimes c\otimes \text{id}_W\big ),
\end{gather}
where $b_{A,B}: a\otimes b\mapsto b\otimes a$ is the trivial braiding in $\text{Vec}$. For $a\in (H'')^*$, $v\in V$, and $w\in W$, this map acts as
\begin{equation}
    r_{(V,c)}(a\otimes v\otimes w) = v_{(i)}\otimes r((S_{(H')^*}(h^{(i)})\triangleright a)\otimes w).
\end{equation}
This linear map defines a $(H'')^*$-representation on $V\otimes W$, since
\begin{eqnarray}
    r_{(V,c)}(ab\otimes v\otimes w) &=& v_{(i)}\otimes r((S_{(H')^*}(h^{(i)})\triangleright ab)\otimes w) 
    \\
    &=& v_{(i)}\otimes r( ((S_{(H')^*}(h^{(i)}))_{(1)}\triangleright a)((S_{(H')^*}(h^{(i)}))_{(2)}\triangleright b))\otimes w)
    \\
    &=& (v_{(i)})_{(j)}\otimes r( ((S_{(H')^*}(h^{(j)}))\triangleright a)((S_{(H')^*}(h^{(i)}))\triangleright b))\otimes w)
\end{eqnarray}
is equal to
\begin{align}
    r_{(V,c)}(a\otimes r_V(b\otimes v\otimes w)) =&  \, \ r_{(V,c)}(a\otimes v_{(i)}\otimes r((S_{(H')^*}(h^{(i)})\triangleright b)\otimes w)) 
    \\
    =& \, \ (v_{(i)})_{(j)}\otimes r(S_{(H')^*}(h^{(j)})\triangleright a \otimes r((S_{(H')^*}(h^{(i)})\triangleright b)\otimes w)
    \\
    =& \, \ (v_{(i)})_{(j)}\otimes r((S_{(H')^*}(h^{(j)})\triangleright a)(S_{(H')^*}(h^{(i)})\triangleright b)\otimes w)
\end{align}
by virtue of the weak action condition (\ref{eq:2cocycletaucondition}).

Moreover, for an intertwiner $f:(W,r)\to (W',r')$, meaning a linear map $f:W\to W'$ satisfying the commutative diagram
\begin{equation}
  \begin{tikzcd}
(H'')^*\otimes W \arrow[rr, "\text{id}_{(H'')^*}\otimes f"] \arrow[dd, "r"'] &  & (H'')^*\otimes W' \arrow[dd, "r'"] \\
                                                                             &  &                                    \\
W \arrow[rr, "f"']                                                           &  & W'                                
\end{tikzcd},
\end{equation}
it is straightforward to check that $f_{{(V,c)}}= \text{id}_V\otimes f: V\otimes W\to V\otimes W'$ is an intertwiner $f_{(V,c)}:(V\otimes W,r_{(V,c)})\to (V\otimes W',r_{(V,c)}')$. Therefore, the assignment
\begin{equation}\label{eq:comoduleaction}
    F_{(V,c)}((W,r)) = (V\otimes W, r_{(V,c)})
\end{equation}
defines a linear endofunctor on $\text{Rep}((H'')^*)$
\begin{equation}
    \rho_{(V,c)}: \text{Rep}((H'')^*) \to \text{Rep}((H'')^*),
\end{equation}
and thus a functor
\begin{equation}
    \rho: \text{Comod}((H')^*)\to \text{End}(\text{Rep}((H'')^*)).
\end{equation}

\subsubsection{Strong monoidal structure}

Now, we define a strong monoidal structure \textit{on the action functor} $\rho$. Generally, this means defining natural isomorphisms
\begin{equation}
    J_{X,Y}: F(X)\otimes F(Y) \xrightarrow{\sim} F(X\otimes Y)
\end{equation}
satisfying the diagram (cf. (\ref{eq:strongmonoidalstr}))
\begin{equation}
  \begin{tikzcd}
(F(X)\otimes F(Y))\otimes F(Z) \arrow[rrr, "{a^{\cal D}_{F(X),F(Y),F(Z)}}"] \arrow[d, "{J_{X,Y}\otimes \text{id}_{F(Z)}}"'] &  &  & F(X)\otimes (F(Y)\otimes F(Z)) \arrow[d, "{\text{id}_{F(X)}\otimes J_{Y,Z}}"] \\
F(X\otimes Y)\otimes F(Z) \arrow[d, "{J_{X\otimes Y,Z}}"']                                                         &  &  & F(X)\otimes F(Y\otimes Z) \arrow[d, "{J_{X,Y\otimes Z}}"]                     \\
F((X\otimes Y)\otimes Z) \arrow[rrr, "{F(a^{\cal C}_{X,Y,Z})}"']                                                            &  &  & F(X\otimes (Y\otimes Z))                                                     
\end{tikzcd}.
\end{equation}

First, let us recall how the tensor product in a comodule category $\text{Comod}(H)$ works. Given a pair of representations $(V,c)$ and $(V',c')$ we can form a new representation on the tensor vector space $V\otimes V'$ along with the linear map
\begin{equation}
    c\otimes c': V\otimes V' \xrightarrow{c\otimes c'} H\otimes V\otimes H\otimes V' \xrightarrow{\text{id}_H\otimes b_{V,H}\otimes \text{id}_{V'}} H\otimes H\otimes V\otimes V'\xrightarrow{\mu\otimes \text{id}_{V\otimes V'}V\otimes V'} H\otimes V\otimes V',
\end{equation}
briefly written as $h^{(i)}\otimes (v\otimes v')_{(i)}= h^{(i)}h^{(j)}\otimes v_{(i)}\otimes v'_{(j)}$. For triples, there are two ways to take the tensor product, namely $(V\otimes V')\otimes V''$ and $V\otimes (V'\otimes V'')$, but on basis elements $v\otimes v'\otimes v''$ the coaction can be written unambiguously as $h^{(i)}h^{(j)}h^{(k)}\otimes v^{(i)}\otimes v'^{(j)}\otimes v''^{(k)}$ since $H$ is associative, so that as long as we work with these explicit expressions we can ignore the associators. Moreover, the category $\text{End}(\text{Rep}((H'')^*))$ is strict, so we can ignore the associators of its tensor product.

For reasons that will be clearer momentarily, we simplify and rewrite the diagram with the arrows in the opposite direction as
\begin{equation}
\begin{tikzcd}
{\rho_{(V,c)}\circ \rho_{(V',c')}\circ \rho_{(V',c')}}                                                                 &  &  &  &  & {\rho_{(V,c)}\circ \rho_{(V'\otimes V'',c'\otimes c'')}} \arrow[lllll, "{\text{id}_{\rho_{(V,c)}}\otimes J_{(V',c'),(V'',c'')}}"']                                           \\
                                                                                                                       &  &  &  &  &                                                                                                                                                                              \\
{\rho_{(V\otimes V,c\otimes c')}\circ \rho_{(V'',c'')}} \arrow[uu, "{J_{(V,c),(V',c')}\otimes \text{id}_{(V'',c'')}}"] &  &  &  &  & {\rho_{(V\otimes V'\otimes V'',c\otimes c'\otimes c'')}} \arrow[uu, "{J_{(V,c),(V'\otimes V'',c'\otimes c'')}}"'] \arrow[lllll, "{J_{(V\otimes V',c\otimes c'),(V'',c'')}}"]
\end{tikzcd},\label{diag:simplifiedmonoidal}
\end{equation}
which encodes the same information since the morphisms $J_{X,Y}$ are isomorphisms.

The image of an $(H'')^*$-module $(W,r)$ under a composition of linear actions, say $\rho_{(V,c)}\circ \rho_{(V',c')}$, is the vector space $V\otimes V'\otimes W$ with the action
\begin{gather}
    (r_{(V',c')})_{(V,c)}(a\otimes v\otimes v'\otimes w) = v_{(i)}\otimes r_{(V',c')}(S(h^{(i)})\triangleright a\otimes v'\otimes w)
    \\
    = v_{(i)}\otimes v'_{(j)}\otimes r(S(h^{(j)})\triangleright (S(h^{(i)})\triangleright a)\otimes w),
\end{gather}
whereas under the functor $\rho_{(V\otimes V',c\otimes c')}$ it is again the vector space $V\otimes V'\otimes W$ but now with the action
\begin{gather}
    r_{(V\otimes V',c\otimes c')}(a\otimes v\otimes v'\otimes w)= (v\otimes v')_{(i)}\otimes r(S(h^{(i)})\triangleright a\otimes w) 
    \\
    = v_{(i)}\otimes v'_{(j)}\otimes r(S(h^{(i)}h^{(j)})\triangleright a\otimes w).
\end{gather}
We define the morphisms
\begin{eqnarray}\label{eq:monoidalstructureforcomoduleaction}
    (J_{(V,c),(V',c')})_{(W,r)}&:& \rho_{(V\otimes V', c\otimes c')}(W,r) \xrightarrow{\sim} \rho_{(V,c)}(\rho_{(V',c')}(W,r))
    \\
    &:&v\otimes v'\otimes w\mapsto v_{(i)}\otimes v'_{(j)}\otimes r(\tau(S(h^{(j)}),S(h^{(i)}))\otimes w),
\end{eqnarray}
whose inverses are given by the convolution inverse of $\tau$
\begin{equation}
    (J_{(V,c),(V',c')})^{-1}_{(W,r)}: v\otimes v'\otimes w\mapsto v_{(i)}\otimes v'_{(j)}\otimes r((\tau(S(h^{(j)}),S(h^{(i)})))^{-1}\otimes w).
\end{equation}
One can check that the linear isomorphisms (\ref{eq:monoidalstructureforcomoduleaction}) are intertwiners by using the twisted module condition (\ref{eq:twistedmodulecondition}) (and the fact that, for a Hopf algebra $H$ with counit $\epsilon$ and antipode $S$, $\epsilon(S(h))=\epsilon(h)$ \cite[Prop. 1.3.1]{Majid}), and that they satisfy the diagram (\ref{diag:simplifiedmonoidal}) by using the 2-cocycle condition (\ref{eq:2cocycletaucondition}).

Hence, we have defined an action of $\text{Comod}((H')^*)$ on $\text{Rep}((H'')^*)$ by linear endomorphisms. 

\subsection{Orbits}

Having described the $\text{Comod}((H')^*)$-module structure on $\text{Rep}((H'')^*)$, we are ready to make the notion of \textit{orbits} precise. For this, let ${\cal I}$ be the set of equivalence classes of irreps of $(H'')^*$. We can define the following relation for any pair $(W,r), (W',r')\in {\cal I}$
\begin{equation}\label{eq:orbitrelation}
    (W,r) \sim (W',r') \iff (W,r) \text{ is a subquotient of } \rho_{(V,c)}(W',r')
\end{equation}
for some $(V,c)\in\text{ob}(\text{Comod}((H')^*))$. For objects $x,y\in\text{ob}({\cal C})$, $x$ is a \textit{subquotient} of $y$ if there exists an object $z\in\text{ob}({\cal C})$ such that $z\hookrightarrow y$ is a subobject (monomorphism) of $y$ and $z\twoheadrightarrow x$ is a quotient (epimorphism) of $z$. 

The relation (\ref{eq:orbitrelation}) is an equivalence relation \cite[Proposition 7.6.6]{EGNO}. By the set of orbits we mean
\begin{equation}\label{eq:setoforbits}
    \mathcal{O}:= \mathcal{I}/\sim,
\end{equation}
the set of equivalence classes. It can be shown \cite[Proposition 7.6.7]{EGNO} that this partition of $\cal I$ extends to the whole $\text{Comod}((H')^*)$-module category, in the sense that $\text{Rep}((H'')^*)$ splits into a direct sum of abelian categories (but clearly not of fusion categories)
\begin{equation}
   \text{Rep}((H'')^*) = \bigoplus_{i\in \mathcal{O}} \mathcal{M}_i, 
\end{equation}
where each $\mathcal{M}_i$ includes the simple objects in the equivalence class $i\in\mathcal{O}$, and is a $\text{Comod}((H')^*)$-module category on its own, so that there is a family of action (tensor) functors
\begin{equation}
    \rho_i:\text{Comod}((H')^*)\to \text{End}(\mathcal{M}_i).
\end{equation}

\subsection{Stabilizers and induced gaugeable algebras}

Finally, for each tensor functor we can consider its kernel $\mathcal{K}_i:=\mathfrak{Ker}_{\rho_i}\subset \text{Comod}((H')^*)$, the full tensor subcategory of objects whose image under $\rho_i$ is isomorphic to a finite direct sum of the identity linear automorphism $\text{id}_{\mathcal{M}_i}:\mathcal{M}_i\to \mathcal{M}_i$. In other words, we are considering the \textit{stabilizer} subcategory, the fusion subcategory whose objects act trivially on the orbit $i$. The restriction of $\rho_i$ to its kernel and corestriction to the category $\langle \text{id}_{\mathcal{M}_i}\rangle = \text{Vec}$ spanned by the identity functor describes a fiber functor on $\mathcal{K}_i$
\begin{equation}
    \rho_i\vert_{\mathcal{K}_i}: \mathcal{K}_i\to \text{Vec},
\end{equation}
giving a Hopf algebra as $\text{End}(\rho_i\vert_{\mathcal{K}_i})$, and thus a symmetric special Frobenius algebra $K_i\in \text{ob}(\mathcal{K}_i)\subset \text{ob}(\text{Comod}((H')^*)$. This is precisely the algebra one gauges for the class $i\in \mathcal{O}$ in the decomposition. 

\subsection{Recovering previous results}

Let us now show how this reproduces the two conjectures presented so far.

\subsubsection{Group-like case}

First, we concentrate on the case of a finite group $G$ with a trivially-acting normal subgroup $N$ and the remaining symmetry group $\Gamma$. We describe this as a sequence of Hopf algebras
\begin{equation}\label{eq:sequenceofdualgroupsalgebras}
    \C^{\Gamma}\to \C^G\to \C^N,
\end{equation}
to give rise via representation categories to the fusion categories
\begin{equation}
    \text{Vec}(N)\to\text{Vec}(G)\to\text{Vec}(\Gamma).
\end{equation}
For the decomposition, we describe a $\text{Vec}(\Gamma)$-action by linear endomorphisms on $\text{Rep}((\C^N)^*)=\text{Rep}(N)$.

Trivially, the dual of the sequence (\ref{eq:sequenceofdualgroupsalgebras}) is the linearization of the group exact sequence
\begin{equation}
    \C N\to \C G\to \C\Gamma,
\end{equation}
which is determined by an action $\triangleright: \C\Gamma\otimes \C N\to \C N$ and a 2-cocycle $\tau: \C\Gamma\otimes \C\Gamma\to \C N$. The action (\ref{eq:comoduleaction}) of a simple object $U_g$ in $\text{Vec}(\Gamma)$ is simply precomposition by the action $g^{-1}\triangleright-$ on the argument of a representation of $N$. This is the same action as \cite[Equation 2.2]{Robbins:2020msp}.

Since this action maps irreps to irreps, the set (\ref{eq:setoforbits}) is simply the set of orbits $\text{Irrep}(N)/\Gamma$.

Now, let $i\in \mathcal{O}=\text{Irrep}(N)/\Gamma$ be an orbit. The module subcategory $\mathcal{M}_i$ is the category spanned by all irreps in the orbit $i$. The kernel $\mathfrak{Ker}_{\rho_i}$ of the action functor
\begin{equation}
    \rho_i: \text{Vec}(\Gamma)\to \mathcal{M}_i
\end{equation}
corresponds to the (normal) subgroup $K_i\leq \Gamma$ acting trivially on the orbit $i$, meaning that each functor $(\rho_i)_{k}: \mathcal{M}_i\to \mathcal{M}_i$ for $k\in K_i$ comes equipped with a natural isomorphism $f_{k}: \text{id}_{\mathcal{M}_i}\xrightarrow{\sim} (\rho_i)_k$. Such natural isomorphisms are intertwiners, linear maps $(f_k)_{(W,r)}: W\to U_k\otimes W$ satisfying the diagram
\begin{equation}
\begin{tikzcd}
\C N\otimes W \arrow[rrr, "{\text{id}_{\C N}\otimes (f_k)_{(W,r)}}"] \arrow[dd, "r"'] &  &  & \C N \otimes U_k\otimes W \arrow[dd, "r_k"] \\
                                                                                      &  &  &                                             \\
W \arrow[rrr, "{(f_k)_{(W,r)}}"']                                                     &  &  & U_k\otimes W                               
\end{tikzcd},
\end{equation}
as required in \cite[Equation 2.9]{Robbins:2020msp}.

Now we describe how to derive the algebra structure on $K_i\in {\cal K}_i$. This is given by the monoidal structure (\ref{eq:monoidalstructureforcomoduleaction}). For this, consider $k,k'\in K_i$. On any irrep $(W,r)$ of $N$, the monoidal structure is the isomorphism
\begin{equation}
    (J_{k,k'})_{(W,r)}: 1_k\otimes 1_{k'}\otimes w\mapsto 1_k \otimes 1_{k'} \otimes r(\tau((k')^{-1},k^{-1})\otimes w).
\end{equation}
However, since $(W,r)$ is an irrep, it satisfies $\text{Hom}((W,r),(W,r))=\C$, so that the composition
\begin{equation}\label{eq:kidiscretetorsion}
\beta^i_{k,k'}:= (f^{-1}_{kk'})_{(W,r)}\circ (J_{k,k'})_{(W,r)}^{-1} \circ (f_k)_{(U_{k'}\otimes W,r_{k'})}\circ (f_{k'})_{(W,r)}: (W,r)\to (W,r)
\end{equation}
is a scalar $\beta_{k,k'}\in \C^{\times}$. This is \cite[Eqn's 2.15,\,2.20]{Robbins:2020msp}.

More explicitly, by choosing the natural isomorphisms $f$ to be
\begin{equation}
    (f_k)_{(W,r)}: w\mapsto 1_g\otimes w,
\end{equation}
the composition (\ref{eq:kidiscretetorsion}) is
\begin{equation}
    w\mapsto 1_{k'}\otimes w\mapsto 1_k\otimes 1_{k'}\otimes w\mapsto 1_{k}\otimes 1_{k'}\otimes r((\tau((k')^{-1},k^{-1}))^{-1}\otimes w)\mapsto r((\tau((k')^{-1},k^{-1}))^{-1}\otimes w),
\end{equation}
and corresponds to scalar multiplication by $\beta^i_{k,k'}\in \C^{\times}$.
Thus, the algebra to gauge is
\begin{gather}
    K_i = \bigoplus_{k\in K_i} U_k,
    \\
    \mu_i(k\otimes k') = \beta_{k,k'}^i kk',
    \\
    \Delta_i(k) = \tfrac{1}{\vert K_i\vert}\sum_{k''\in K_i}(\beta_{k'',(k'')^{-1}k}^i)^{-1} \, k''\otimes (k'')^{-1}k.
\end{gather}
The partial trace $Z_{k,k'}$ therefore has a coefficient
\begin{equation}
    (\mu_i)_{k,k'}^{kk'} = \delta_{kk',k'k} \frac{\beta^i_{k,k'}}{\beta^i_{k',k}}.
\end{equation}

We can specialize further to $N$ abelian, in which case all irreducible representations are one-dimensional, so that by definition for any irrep $(W,r)$ the action $r((\tau((k')^{-1},k^{-1}))^{-1}\otimes w)$ is multiplication by the scalar $\rho((\tau((k')^{-1},k^{-1}))^{-1})\in U(1)$ when regarding the representation map $r:\C N \otimes W\to W$ as a group homomorphism $\rho:N\to U(1)$. Denoting the composition $\omega:=\rho\circ \tau: K_i\times K_i\to U(1)$, the phase for the partial trace $Z_{k,k'}$ in this case becomes
\begin{gather}
  \frac{\omega(k^{-1},(k')^{-1})}{\omega((k')^{-1},k^{-1})}.
\end{gather}
This phase is identical to the usual expression $\epsilon(g,h)=\frac{\omega(g,h)}{\omega(h,g)}$, as one can deduce using the identities $\epsilon(gh,k)=\epsilon(g,k)\epsilon(h,k)$ and $\epsilon(g,1)=1$ \cite{Vafa:1986wx} (which are a consequence of the normalized 2-cocycle identity $\omega(g,h)$ satisfies)
\begin{gather}
    1= \epsilon(gg^{-1},hh^{-1}) = \epsilon(g,h)\epsilon(g,h^{-1})\epsilon(g^{-1},h)\epsilon(g^{-1},h^{-1}),\label{eq:epsilon1}
    \\
    \frac{\epsilon(g^{-1},h^{-1})}{\epsilon(g,h)}= \big(\epsilon(h^{-1},g)\epsilon(h,g) \big) \big(\epsilon(h,g^{-1})\epsilon(h,g) \big) = (\epsilon(1,g))(\epsilon(h,1)) = 1. \label{eq:epsilon2}
\end{gather}

\subsubsection{Hopf abelian extension case}

We can, on the other hand, obtain the algebra structure derived in Section~\ref{sec:pfuncdecomposition} in this way. We remind the reader the setting is an exact sequence
\begin{equation}
\text{Rep}(G)\to \text{Rep}(H)\to \text{Vec}(\Gamma),
\end{equation}
so that we must exhibit a $\text{Vec}(\Gamma)$-module structure on $\text{Rep}((\C G)^*)=\text{Vec}(G)$. On the simple objects, this comes from the right action by permutation (\ref{eq:gammaactong}), turned into a left-action by using the antipode, meaning
\begin{equation}
    \rho_{U_g} (U_x) \cong U_{x\triangleleft g^{-1}}.
\end{equation}

The equivalence classes $\mathcal{O}$ (\ref{eq:setoforbits}) are simply the orbits of $G$ as a $\Gamma$-set.

On a given orbit $[x]\in \mathcal{O}$, the kernel of the restricted action $\rho_i$ consists of (the linear completion of) the subgroup $K_i\leq \Gamma$ fixing the orbit $[x]$. As before, each of the functors $(\rho_i)_{U_k}: \mathcal{M}_i\to \mathcal{M}_i$ comes with a natural isomorphism $f_k: \text{id}_{\cal M_i}\xrightarrow{\sim} (\rho_i)_k$. Since the action of $k$ necessarily sends simples to simples, we can again define scalars
\begin{equation}\label{eq:kidiscretetorsion-abext}
\beta^i_{k,k'}:= (f^{-1}_{kk'})_{U_x}\circ (J_{k,k'})_{U_x}^{-1} \circ (f_k)_{U_{x\triangleleft (k')^{-1}}}\circ (f_{k'})_{U_x}: U_x\to U_x,
\end{equation}
which after choosing isomorphisms $f_k: 1_x\mapsto 1_g\otimes 1_x$,
we get
\begin{equation}
    \beta^i_{k,k'}= (\tau_x((k')^{-1},k^{-1})^{-1}.
\end{equation}
Therefore, the algebra structure is given by
\begin{gather}
    K_i = \bigoplus_{k\in K_i} U_k,
    \\
    \mu_i(k\otimes k') = \beta_{k,k'}^i kk',
    \\
    \Delta_i(k) = \tfrac{1}{\vert K_i\vert}\sum_{k''\in K_i}(\beta_{k'',(k'')^{-1}k}^i)^{-1} \, k''\otimes (k'')^{-1}k,
\end{gather}
much as in the previous case.

Therefore, the phases for each twisted sector $Z_{k,k'}$ are
\begin{equation}
    \delta_{kk',k'k} \, \frac{\tau_x(k^{-1},(k')^{-1})}{\tau_x((k')^{-1},k^{-1})}.
\end{equation}
Given that for each $x\in [x]$, the function $\tau_x$ is a 2-cocycle valued on $K_i$ valued in $U(1)$, with trivial action as $K_i$ fixes $x$, then the derivation (\ref{eq:epsilon1})-(\ref{eq:epsilon2}) follows identically. This implies that the phase for $Z_{k,k'}$ obtained previously and obtained here match
\begin{equation}
    \frac{\tau_x(k,k')}{\tau_x(k',k)}= \frac{\tau_x(k^{-1},(k')^{-1})}{\tau_x((k')^{-1},k^{-1})},
\end{equation}
so that we obtain the same partition function.

\section{Conclusion}
In this paper, we conjectured the decomposition of a theory $\cal T$ with a $\text{Rep}(H)$ symmetry, for $H$ a finite-dimensional semisimple Hopf algebra, with a trivially-acting subcategory $\text{Rep}(G)$ and $\text{Vec}(\Gamma)$ the remaining symmetry, for $G,\Gamma$ finite groups. Mathematically, this was described by abelian extensions of Hopf algebras, extensions of the form 
\begin{equation}
    \C^{\Gamma}\to H\to \C G,
\end{equation}
which give rise to exact sequences of fusion categories
\begin{equation}
    \text{Rep}(G)\to\text{Rep}(H)\to \text{Vec}(\Gamma).
\end{equation}
We explicitly derived the decomposition of the $\text{Rep}(H)$-gauged partition function, and computed the topological projection responsible for this decomposition.

Motivated by the observation that, in the abelian extension case, the algebra extension information is not relevant for the decomposed partition function, combined with alternative constructions of gaugeable algebras which are independent of the algebra structure of $H$, we commented on a plausible decomposition conjecture and provided a general construction which computes the set of universes and their corresponding non-invertible orbifolds (gaugeable algebras). This outlines a decomposition conjecture for a $\text{Rep}(H)$ symmetry category with a $\text{Rep}(H'')$ trivially-acting subcategory. 

We hope to extend in future work the explicit analysis of Section~\ref{sec:pfuncdecomposition} to more general Hopf algebra extensions, as well as the computation of more examples. Also, as indicated in Section~\ref{ssec:projection-operators} cf. Footnote~\ref{footnote:mixed-anomaly-phases}, it would be interesting to provide a more systematic explanation of the projective (``mixed-anomaly'') phases discussed in \cite{Robbins:2022wlr}.

Furthermore, one should note that exact sequences of fusion categories, not necessarily of the form $\text{Rep}(H)$ for $H$ a Hopf algebra, are very generally classified by a \textit{Hopf monad} on what is understood as the remaining symmetry category. It has been suggested in \cite{CZW19} that Hopf monads may play a role in the description of generalized symmetries in two dimensions, and decomposition in this generality seems like a natural application of this. We hope to address this in future work.

\section{Data availability}

We do not analyse or generate any datasets, because our work
proceeds within a theoretical and mathematical approach.

\section{Declarations}

The author declares no competing interests.

\appendix

\section{Notes on exact sequences and extensions}\label{app:notes}

Here, we provide the more technical part of the mathematical definitions presented in Section~\ref{sec:mathbackground}, following \cite{BN11}.

\subsection{Exact sequences of fusion categories}\label{sapp:exactsequencesoffusion}
An exact sequence of fusion categories is a diagram of fusion categories and functors of the form
\begin{equation}
    {\cal K} \xrightarrow{\imath} {\cal C}\xrightarrow{F} {\cal D}.
\end{equation}
These fusion categories and functors need to satisfy specific conditions, which we summarize as follows
\begin{enumerate}
    \item The functors $\imath:{\cal K\to C}$ and $F:{\cal C\to D}$ are tensor functors,
    \item $F$ is a normal, dominant tensor functor,
    \item $\imath$ is a full embedding,
    \item the image $\imath({\cal K})\subset {\cal C}$ is tensor equivalent to the kernel $\mathfrak{Ker}_F$ of $F$.
\end{enumerate}

Let us elaborate on these requirements. First, a \textit{tensor} functor $F:{\cal C\to D}$ a $\C$-linear \textit{strong} monoidal functor \cite{EGNO}, which in particular implies that if comes equipped with natural \textit{iso}morphisms
\begin{equation}
    J_{X,Y}: F(X)\otimes F(Y)\xrightarrow{\sim} F(X\otimes Y),
\end{equation}
satisfying the commutative diagram (\ref{eq:strongmonoidalstr})
\begin{equation}\label{eq:strongmonoidalstr}
\begin{tikzcd}
(F(X)\otimes F(Y))\otimes F(Z) \arrow[rrr, "{a^{\cal D}_{F(X),F(Y),F(Z)}}"] \arrow[d, "{J_{X,Y}\otimes \text{id}_{F(Z)}}"'] &  &  & F(X)\otimes (F(Y)\otimes F(Z)) \arrow[d, "{\text{id}_{F(X)}\otimes J_{Y,Z}}"] \\
F(X\otimes Y)\otimes F(Z) \arrow[d, "{J_{X\otimes Y,Z}}"']                                                         &  &  & F(X)\otimes F(Y\otimes Z) \arrow[d, "{J_{X,Y\otimes Z}}"]                     \\
F((X\otimes Y)\otimes Z) \arrow[rrr, "{F(a^{\cal C}_{X,Y,Z})}"']                                                            &  &  & F(X\otimes (Y\otimes Z))                                                     
\end{tikzcd},
\end{equation}
for all $X,Y,Z\in \text{ob}({\cal C})$. Here, $a^{\cal C}$ and $a^{\cal D}$ refer to the associators in $\cal C,D$, respectively.

Second, a tensor functor $F:{\cal C\to D}$ is \textit{dominant} if, for every object $d\in\text{ob}({\cal D})$, there exists an object $c\in \text{ob}({\cal C})$ and a monomorphism $d\hookrightarrow F(c)$, meaning $d$ is a \textit{subobject} of $F(c)$. The tensor functor $F$ is, on the other hand, called \textit{normal} if given an object $c\in\text{ob}({\cal C})$ there exists an object $c_0\in\text{ob}({\cal C})$ such that $F(c_0)$ is the largest \textit{trivial} subobject of $F(c)$. Here, trivial means that it is isomorphic to the direct sum of a finite number of copies of the monoidal unit $\mathbbm{1}_{\cal D}\in\text{ob}({\cal D})$.

Third, a functor $\imath: {\cal K\to C}$ is called a \textit{full embedding} if it is injective on the objects, and bijective on the hom-sets $\imath(\text{Hom}_{\cal C}(X,Y))\cong \text{Hom}_{\cal D}(\imath(X),\imath(Y))$.

Finally, we require that the fusion subcategories $\imath({\cal K}), \mathfrak{Ker}_F\subset {\cal C}$ are equivalent as tensor categories, meaning there exists an equivalence of categories for which the functors and natural transformations are tensor functors and monoidal natural transformations. Here, the fusion subcategory $\mathfrak{Ker}_F$ is the full subcategory spanned by all objects $c\in\text{ob}({\cal C})$ such that their image under $F$ is a trivial object in $\cal D$. For this reason, this subcategory is referred to as the \textit{kernel} of $F$.

Several examples of these sequences come from group exact sequences. Given a group exact sequence
\begin{equation}
    1\to N\to \Gamma\to G\to 1,
\end{equation}
this gives rise to an exact sequence of fusion categories of graded vector spaces
\begin{equation}
    \text{Vec}(N)\to \text{Vec}(\Gamma)\to \text{Vec}(G),
\end{equation}
as well as an exact sequence of group representations
\begin{equation}
    \text{Rep}(G)\to \text{Rep}(\Gamma)\to \text{Rep}(N)
\end{equation}
(note the change of direction). These two kinds of exact sequences play a central role in \cite{Perez-Lona:2023djo,Perez-Lona:2024sds}.

\subsection{Exact sequences of Hopf algebras}\label{sapp:hopfexact}
We spell out the details of Hopf algebra exact sequences and their defining extension data. 

\subsubsection{General extensions}\label{sapp:generalextensions}
An exact sequence, equivalently an \textit{extension}, of Hopf algebras is a diagram of the form
\begin{equation}\label{eq:exactseqhopf}
    H' \xrightarrow{i} H \xrightarrow{\pi} H'', 
\end{equation}
where $i,\pi$ are Hopf algebra homomorphisms, such that
\begin{enumerate}
    \item $i$ is injective, and $\pi$ is surjective,
    \item $i(H')=\{h\in H\vert (\pi\otimes\text{id}_H)\Delta(h)= 1\otimes h\}$,
    \item $\text{ker}\,\pi=\{ab\in H\vert a\in H, \ b\in \text{ker}(\epsilon\vert_{i(H')})\}$.
\end{enumerate}
The set $\text{ker}(\epsilon\vert_{i(H')})$ is the kernel of the restricted counit $\epsilon\vert_{i(H')}:i(H')\to \C$, and is also known as the \textit{augmentation ideal} of $i(H')$.

Sequence of this form (\ref{eq:exactseqhopf}), as in the case of groups, are classified by \textit{extension data} involving only the Hopf algebras $H'$ and $H''$. The extension data consists of four linear maps $(\triangleright, \rho,\sigma,\tau)$: the \textit{weak action} $\triangleright: H''\otimes H'\to H'$, the \textit{weak coaction} $\rho: H''\to H''\otimes H'$, the \textit{2-cocycle} $\sigma: H''\otimes H''\to H'$, and the \textit{dual 2-cocycle} $\tau:H''\to H'\otimes H'$. These linear maps are required to satisfy an array of compatibility conditions, listed in \cite[Theorem 2.20]{AD95}). Succinctly, the maps $(\triangleright,\sigma)$ determine the \textit{algebra} structure on $H$, whereas the maps $(\rho,\tau)$ determine the \textit{coalgebra} structure.

In this paper, for more general Hopf algebra extensions, we will only make explicit use of the identities satisfied by the maps $(\triangleright,\sigma)$:
\begin{enumerate}
    \item normalized 2-cocycle condition
    \begin{gather}
    \sigma (1,h)= \sigma(h,1)= \epsilon(h)\,1,
    \\
    (h_{(1)}\triangleright \sigma(l_{(1)},m_{(1)}) \sigma(h_{(2)},l_{(2)}m_{(2)}) = \sigma(h_{(1)},l_{(1)})\sigma(h_{(2)}l_{(2)},m), 
\end{gather}
\item weak action condition
\begin{gather}
    h\triangleright (ab) = (h_{(1)}\triangleright a)(h_{(2)}\triangleright b),
    \\
    h\triangleright 1 = \epsilon(h) \, 1,
    \\
    1\triangleright a = a.
\end{gather}
\item twisted module condition

\begin{equation}
    (h_{(1)}\triangleright (l_{(1)}\triangleright a))\tau(h_{(2)},l_{(2)}) = \sigma(h_{(1)},l_{(1)})(h_{(2)}l_{(2)}\triangleright a).
\end{equation}
\end{enumerate}
In stating these identities, we are making use of \textit{Sweedler's notation} (see Appendix~\ref{app:sweedler}).

As a quick example, we can specialize to group algebras with coalgebra extension information $(\rho,\tau)$ that is all trivial:
\begin{equation}
    \C N \to \C \Gamma\to \C G.
\end{equation}
Since the coalgebra structure of group algebras themselves is trivial, in the sense that $\Delta(g) = g\otimes g$ and $\epsilon(g)=1$, then the extension information reduces to the familiar 2-cocycle with nontrivial action. The group algebra cocycles and action are the linear extension of cocycles and actions on the groups. Indeed, for $h,l,m\in G$ and $a,b\in \Gamma$, the identities become
 \begin{gather}
    \sigma (1,h)= \sigma(h,1)= 1,
    \\
    (h\triangleright \sigma(l,m)) \sigma(h,lm) = \sigma(h,l)\sigma(hl,m), 
\end{gather}
\begin{gather}
    h\triangleright (ab) = (h\triangleright a)(h\triangleright b),
    \\
    h\triangleright 1 = \, 1,
    \\
    1\triangleright a = a,
\end{gather}
where in particular the action $\triangleright:G\times\Gamma\to \Gamma$ becomes a $G$-action by homomorphisms on $\Gamma$.

\subsubsection{Abelian extensions}\label{sapp:abelian}

Specializing to abelian extensions
\begin{equation}\label{eq:abextensionapp}
    \C^{\Gamma}\to H\to \C G,
\end{equation}
 the extension data can be rewritten as follows \cite{Nat20} (cf. \cite{Tak81,Mat02}):
\begin{itemize}
\item A \textit{matched pair} \cite{Kac68} of groups $(G,\Gamma,\triangleright,\triangleleft)$: a pair of finite groups $G,\Gamma$ along with maps of sets
\begin{eqnarray}
    \triangleleft:& \Gamma\times G\to \Gamma,
    \\
    \triangleright:& \Gamma \times G \to G, 
\end{eqnarray}
satisfying the conditions
\begin{eqnarray}
    s\triangleright gh = (s \triangleright g)((s \triangleleft g) \triangleright h), \label{eq:trianglerightidentity}
    \\
    st \triangleleft g = (s \triangleleft (t \triangleright g))(t \triangleleft g), \label{eq:triangleleftifentity}
\end{eqnarray}
for any $s, t \in \Gamma$ and $ g, h \in G$ (this encompasses the action and coaction),
\item a pair of 2-cocycles $\sigma: G\times G\to (\C^{\times})^{\Gamma}$, $\tau: \Gamma\times\Gamma\to (\C^{\times})^G$ satisfying
\begin{gather}
    \sigma_{s\triangleleft g}(h,l)\sigma_s(g,hl) = \sigma_s(g,h)\sigma_s(gh,l), \label{eq:sigma2-cocycle}
    \\
    \sigma_s(1,g)=\sigma_s(g,1)=1, \label{eq:sigmanormalization}
    \\
    \tau_g(st,u)\tau_{u\triangleright g}(s,t)=\tau_g(t,u)\tau_g(s,tu),
     \label{eq:tau2cocycle} \\
    \tau_g(1,s)=\tau_g(s,1)=1,
    \\
    \sigma_{st}(g,h)\tau_{gh}(s,t)=\sigma_s(t\triangleright g, (t\triangleleft g)\triangleright h)\sigma_t(g,h)\tau_g(s,t)\tau_h(s\triangleleft (t\triangleright g),t\triangleleft g),
    \\
    \sigma_1 (g,h) = \tau_1 (s,t) = 1, \label{eq:2cocyclesevaluatedatone}
\end{gather}
using the notation $\sigma_s(g,h)=\sigma(g,h)(s)$ (and similarly for $\tau_g(t,u)$), for $g,h,l\in G$ and $s,t,u\in \Gamma$.
\end{itemize}

The vector space $\C^{\Gamma}\otimes \C G$, with basis elements $v_g\# x$ for $x\in G$ the usual basis of a group algebra, and $v_g$ the dual basis of the basis element given by $g\in \Gamma$, has the Hopf algebra structure \cite{AD95,Mat02,AN03} (cf. Eq's (\ref{eq:abextunit})-(\ref{eq:abantipode})),
\begin{gather}
    u: 1\mapsto \sum_{g\in\Gamma} v_g\otimes 1,\\
\mu: (v_g\# x)\otimes (v_h\# y) \mapsto  \delta_{g\triangleleft x,h}\sigma_g(x,y)v_g\# xy, \\
    \epsilon: (v_g\# x) \mapsto \delta_{1,g}, \\
    \Delta: (v_g\# x) \mapsto \sum_{t\in \Gamma} (\tau_x(gt^{-1},t)v_{gt^{-1}}\# t\triangleright x)\otimes (v_t\# x),
\end{gather}
\begin{equation}
    S(v_g\# x) = \big(\sigma_{(g \triangleleft x)^{-1}}((g\triangleright x)^{-1}, g\triangleright x)\big)^{-1} \, \big(\tau_x(g^{-1}, g)\big)^{-1}\, v_{(g\triangleleft x)^{-1}} \# (g\triangleright x)^{-1}.
\end{equation}

Finally, let us compute an integral and a cointegral of $H$. An integral \cite{LS69} for $H$ is an element $\Lambda\in H$ such that for every $h\in H$ the product is
\begin{equation}
    h\Lambda = \epsilon(h)\Lambda.
\end{equation}
A cointegral is a linear function $\lambda$ of $H$, an element $\lambda\in H^*$, such that
\begin{equation}
    (\lambda\otimes\text{Id}_H)(\Delta(h)) = \lambda(h)1_H.
\end{equation}

One can check that an integral $\Lambda\in H$ is
\begin{equation}\label{eq:hopfintegralapp}
    \Lambda = v_1\# \left(\sum_{x\in G}x\right),
\end{equation}
since $\triangleleft$ is an action by permutation that $1\triangleleft x=1$, so that
\begin{eqnarray*}
  (v_g\# x)\left(v_1\# \left(\sum_{y\in G}y\right)\right) &=& \sum_{y\in G} \delta_{g\triangleleft x,1} \sigma_g(x,y) v_g \# xy , 
  \\
  &=& \delta_{g,1}\sum_{y\in G} v_1\# xy ,
  \\
  &=& \epsilon(v_g\# x)\left(v_1\# \left(\sum_{y\in G}y\right)\right).
\end{eqnarray*}
A cointegral $\lambda$ of $H$ is the function
\begin{equation}
    \lambda=\left(\sum_{g\in\Gamma} g\right)\#v_1,
\end{equation}
as one can verify
\begin{eqnarray*}
    \left(\sum_{k\in\Gamma}k\# v_1 \otimes \text{Id}_H\right)(\Delta(v_g\# x)) &=&  \left(\sum_{k\in\Gamma}k\# v_1 \otimes \text{Id}_H\right)\left(\sum_{t\in\Gamma} (\tau_x(gt^{-1},t)v_{gt^{-1}}\# t \triangleright x)\otimes (v_t\# x)\right),
    \\
    &=& \delta_{1,x} \sum_{t\in \Gamma} v_{t}\# 1,
    \\
    &=& \delta_{1,x} \, 1_H = \lambda(v_g\# x) 1_H.
\end{eqnarray*}

\section{Sweedler's notation}\label{app:sweedler}
Throughout the paper, we make use of Sweedler's notation. Given a coalgebra $H$ with comultiplication $\Delta$, the coproduct of any element $h\in H$ is represented as
\begin{equation}
    \Delta(h) = h_{(1)}\otimes h_{(2)},
\end{equation}
where a sum over indices is assumed. Each term $h_{(1)},h_{(2)}\in H$ is an element of the coalgebra. In particular, in this notation, coassociativity of $\Delta$ can be represented as
\begin{equation}
    (h_{(1)})_{(1)}\otimes (h_{(1)})_{(2)} \otimes h_{(2)} = h_{(1)}\otimes (h_{(2)})_{(1)}\otimes (h_{(2)})_{(2)} = h_{(1)}\otimes h_{(2)}\otimes h_{(3)}.
\end{equation}

Sweedler's notation also applies for comodules. Given a vector space $V$ with a coaction $\rho: V\to V\otimes H$ by a coalgebra $H$, the coaction on an element $v\in V$ is represented as the implicit sum
\begin{equation}
    \rho(c)= c_{(i)}\otimes h^{(i)}.
\end{equation}

\end{document}